\documentclass[fleqn,usenatbib]{mnras}

\usepackage{newtxtext,newtxmath}

\usepackage[T1]{fontenc}
\usepackage{ae,aecompl}

\usepackage{graphicx}	
\usepackage{amsmath}	
\usepackage{amssymb}	
\usepackage{mathrsfs}
\usepackage{xcolor}

\newcommand{\msun}{\mathrm{M}_\odot}
\newcommand{\mstar}{M_\star}
\newcommand{\subfind}{\textsc{Subfind}}
\newcommand{\modtext}{\color{black}}

\title[SMF and density profile evolution of Hydrangea simulations at 0<z<1.5]{The stellar mass function and evolution of the density profile of galaxy clusters from the Hydrangea simulations at $0<z<1.5$}

\author[S. L. Ahad et al.]{Syeda Lammim Ahad,$^{1}$\thanks{E-mail: ahad@strw.leidenuniv.nl}
Yannick M. Bah\'{e},$^{1}$
Henk Hoekstra,$^{1}$
\newauthor Remco F. J. van der Burg,$^{2}$
Adam Muzzin$^{3}$
\\
$^{1}$ Leiden Observatory, Leiden University, P.O. Box 9513, 2300 RA Leiden, The Netherlands\\
$^{2}$European Southern Observatory, Karl-Schwarzschild-Str. 2, 85748, Garching, Germany\\
$^{3}$Department of Physics and Astronomy, York University, 4700, Keele Street, Toronto, ON, MJ3 1P3, Canada
}

\date{Accepted XXX. Received YYY; in original form ZZZ}

\pubyear{2020}

\begin{document}
\label{firstpage}
\pagerange{\pageref{firstpage}--\pageref{lastpage}}
\maketitle

\begin{abstract}
Galaxy clusters are excellent probes to study the effect of environment on galaxy formation and evolution. Along with high-quality observational data, accurate cosmological simulations are required to improve our  understanding of galaxy evolution in these systems. In this work, we compare state-of-the-art observational data of massive galaxy clusters ($>10^{14}\, \textrm{M}_{\odot}$) at different redshifts ($0<z<1.5$) with predictions from the Hydrangea suite of cosmological hydrodynamic simulations of 24 massive galaxy clusters ($>10^{14}\, \textrm{M}_{\odot}$ at $z=0$). We compare three fundamental observables of galaxy clusters: the total stellar mass to halo mass ratio, the stellar mass function (SMF), and the radial mass density profile of the cluster galaxies. In the first two of these, the simulations agree well with the observations, albeit with a slightly too high abundance of $M_\star \lesssim 10^{10} \,\msun$ galaxies at $z \gtrsim 1$. The NFW concentrations of cluster galaxies increase with redshift, in contrast to the decreasing dark matter halo concentrations. This previously observed behaviour is therefore due to a qualitatively different assembly of the smooth DM halo compared to the satellite population. Quantitatively, we however find a discrepancy in that the simulations predict higher stellar concentrations than observed at lower redshifts ($z<0.3$), by a factor of $\approx$2. This may be due to selection bias in the simulations, or stem from shortcomings in the build-up and stripping of their inner satellite halo.
\end{abstract}




\section{Introduction}

During the last few decades, our understanding of the formation and evolution of galaxies and the large scale structures they are part of has increased significantly. The $\Lambda$ cold dark matter ($\Lambda$CDM) model has been extremely successful in describing the gravity-driven growth of structures from dwarf galaxies to super-clusters, based on a Universe dominated by dark matter (DM) and, at low redshift, dark energy. One of the most important questions, however, still remains: how exactly did baryonic structures such as galaxies form and grow in this Universe? A particularly useful setting to explore this question is in clusters of galaxies -- the largest gravitationally bound structures in the Universe. Harbouring thousands of galaxies in a relatively small volume, they are akin to cosmic laboratories to study the impact of galaxy interactions and environment on galaxy formation. For example, many observational studies have concluded that at a fixed stellar mass, galaxies in denser environments (especially in groups or clusters) are more likely to be elliptical in morphology \citep{dressler1980ApJ...236..351D}, lack or have low levels of recent or ongoing star formation \citep{dressler1980ApJ...236..351D, balogh1999c, kauffmann2004environmental, weinmann2006properties, blanton2005relationship, Peng_2010, wetzel2012galaxy, woo2012dependence}, and are devoid of atomic hydrogen \citep[e.g.][]{ giovanelli1985gas, fabello2012MNRAS, hess2013AJ,Odekon_et_al_2016, Brown_et_al_2017} in comparison with galaxies in more isolated, ``field'', environments.

The arguably most fundamental observable property of galaxies and clusters is their total stellar mass and the way in which this is distributed over differently massive galaxies -- i.e. the stellar mass function (SMF) -- and galaxies at different cluster-centric radii. The SMF is the product of star formation, galaxy mergers, and stripping, so that its shape and temporal evolution provide an observationally accessible tracer of galaxy evolution, and of the role of environment in this process. In simulations, model adjustments typically manifest themselves as pronounced differences in the predicted field SMF \citep[e.g.][]{crain2015eagle}, which has therefore emerged as a prime calibration diagnostic for cosmological hydrodynamic simulations of representative volumes such as EAGLE \citep{schaye2014eagle} or IllustrisTNG \citep{illustristng2017}. Furthermore, several observational studies have reported differences between the satellite and field stellar mass functions \citep[e.g.][]{Yang_et_al_2009,Wang_White_2012,Vulcani_et_al_2014}, so that the cluster SMF also provides a valuable test for the validity of theoretical models.

Another important indicator of the growth of cluster haloes and the formation and evolution of galaxies within them are the radial density profiles of total matter and galaxies in the clusters. Simulations robustly predict that the density profile of the dominant CDM component is well described by the Navarro-Frenk-White (NFW) profile \citep{nfw1996,navarro1997universal,schaller2015effect}, which can be parameterised in terms of the halo concentration and mass. A clear prediction from these simulations is that the concentration of cluster haloes increases with cosmic time \citep[e.g.][]{duffy2008dark,Munoz-Cuartas_et_al_2011} -- in other words, the dark matter in clusters with fixed halo mass (and also of individual evolving clusters) was less concentrated at high redshift than it is now.

Measuring these profiles observationally is challenging, however, because indirect detection methods such as gravitational lensing have to be used \citep[e.g.][]{williams2018hff,Mahler_2019,Gonzalez2020buffalo}. The (projected) satellite stellar mass density profile, on the other hand, can be measured quite clearly and robustly from observations\footnote{Most observational work has focused on the profile of the stellar mass in satellites, i.e. the stellar-mass-weighted satellite halo profile. The diffuse stellar halo of the cluster itself, the intra-cluster light (ICL) has only recently become accessible observationally beyond the very nearby Universe (e.g.~\citealt{Mihos_et_al_2005,Montes_Trujillo_2018,Zhang_et_al_2019}). We do not consider the ICL in this paper, but refer the interested reader to \citet{Alonso_Asensio_et_al_2020} and \citet{Deason_et_al_2020} for recent analyses of its low-redshift properties and radial profiles in Hydrangea.}. Since both dark matter and stars are, essentially, collisionless fluids on cluster scales, one might expect to observe a similar concentration evolution for these stellar profiles as what is predicted for DM. Surprisingly, recent observational studies have revealed a rather different picture: stellar concentrations were much higher than DM predictions at redshift $z \approx 1$ \citep{Muzzin_2007}, but then became less concentrated over cosmic time, and at $z \approx 0$ the stellar concentration is a factor of $\approx$2 lower \citep{Budzynski2012MNRAS,van2015evidence} than what simulations predict for DM.

While it may be tempting to interpret this discrepancy as a failure of $\Lambda$CDM, a more straightforward solution is that the DM and stellar components of galaxy clusters are built up differently over time. It is therefore important to test this interpretation by analysing the evolution of stellar density profiles in cosmological hydrodynamic simulations, which self-consistently model the formation and evolution of stellar mass. Comparing the predicted stellar density profiles to both observations and the simulated DM density profiles can then reveal whether the observed decrease in stellar concentration over time is indeed explained by astrophysical effects, or hints at a need to modify the $\Lambda$CDM paradigm. 

For a simulation to be of use in this experiment, it must as a minimum requirement reproduce the observed field SMF and its evolution ; the simulation resolution must also be high enough that cluster-specific processes such as star formation quenching and stellar stripping can be modelled reliably. The EAGLE simulations satisfy both of these needs: the field SMF is reproduced both at $z \approx 0$, where it was used as a calibration diagnostic \citep{crain2015eagle,schaye2014eagle} and at higher redshift \citep{Furlong2015eaglesmf}, and at a baryon mass resolution of $\approx 2\times 10^6\, \msun$, Milky Way like galaxies are resolved by $\gtrsim10^4$ star particles each. Comparably successful projects include IllustrisTNG \citep{pillepich2018illustris}, Horizon-AGN \citep{dubois-horizon}, and Simba \citep{Dave_et_al_2019}.

However, galaxy clusters inhabit only a small volume fraction of the Universe, so that they are not generally represented (at least not in statistically significant numbers) in simulations comparable to the resolution of EAGLE, which are typically limited to volumes of $\leq$ (100 cMpc)$^3$. Larger-volume simulations, such as  IllustrisTNG-300 \citep{pillepich2018illustris} or BAHAMAS \citep{McCarthy2017Bahama} -- which contain galaxy clusters in sufficient numbers -- in turn still lack the resolution to model individual cluster galaxies in detail. This motivates the use of zoom-in cluster simulations, where only a carefully chosen region within a large volume (such as the site of a massive galaxy cluster) is simulated at high resolution.

In this work, we perform a detailed analysis of one set of such zoom-in simulations, the Hydrangea cluster suite \citep{bahe2017hydrangea,barnes2017cluster}, to gain insight into the evolution of the stellar component of galaxy clusters over cosmic time. Various predictions from these simulations in the local Universe have previously been compared to observations and found to be broadly realistic, including the stellar mass function, quenched fractions \citep{bahe2017hydrangea}, total gas and stellar content \citep{barnes2017cluster}. It is far from guaranteed, however, that these agreements will also hold at higher redshift, especially considering that the simulation model was calibrated at $z \approx 0$ (albeit on the field, not clusters). We therefore perform careful tests of simulation predictions at higher redshift, up to $z\approx1.4$, by comparing them with state-of-the-art observational data. Specifically, we analyse the total stellar content, the satellite stellar mass function, and the radial distribution of stellar mass within the cluster; we then use the latter as a basis to interpret the difference between the observed evolution of stellar concentration and that predicted for DM from $N$-body simulations over cosmic time. 

The remainder of this paper is structured as follows. In Section \ref{simdata}, we describe the key aspects of the Hydrangea simulations, and test their predicted evolution of the total stellar content. In Section \ref{obsdata}, we outline the observational data sets that we have used to test the simulations. The predicted SMF is compared to observations at various redshifts in Section \ref{smfdis}, followed by a comparison of the predicted and observed stellar density profiles in Section \ref{nfwdis}. Finally, in Section \ref{sumandconc}, we summarize our findings and present our conclusions. 

\section{Simulated Data}
\label{simdata}
\subsection{The Hydrangea Simulation Suite}
\label{hydrangea}
The Hydrangea simulations \citep{bahe2017hydrangea, barnes2017cluster} are a suite of high-resolution cosmological hydrodynamic zoom-in simulation of $24$ massive galaxy clusters. They are part of the 30 clusters of the `C-EAGLE' project, and were chosen from a low-resolution N-body (DM only) parent simulation \citep{barnes2017redshift} of a (3200~cMpc$)^3$ volume\footnote{Throughout this paper, we use the prefixes `c' and `p'  to refer to comoving and proper quantities, respectively: `cMpc', for example, denotes `co-moving Mpc'.}. Each of the high-resolution simulation regions is centred on a massive cluster with $M_{200c}$ in the range $10^{14.0}$--$10^{15.4}\,\msun$ at $z = 0$\footnote{$M_{200c}$ refers to the mass enclosed within a sphere centred at the potential minimum of the cluster radius $r_{200c}$, within which the average density of matter equals 200 times the critical density.}. The particle mass resolution is $m_\textrm{baryon} =1.81 \times 10^6\,\msun$ for baryons and $m_\textrm{DM} = 9.7 \times 10^6\,\msun$ for dark matter, respectively\footnote{\citet{Ludlow_et_al_2019} have recently shown that this (common practice) use of lower resolution for the DM component leads to an artificial transfer of energy from DM to stars, which can puff up the central regions of galaxies. We do not expect this to have an effect on our study, since the concentrations we measure in Section \ref{denprofsec} are derived from the (unaffected) radii of galaxies within the cluster, rather than the distribution of stars within them.}; the gravitational softening length is $\epsilon=0.7$ physical kpc (pkpc) at $z<2.8$. The high-resolution simulation regions include the large scale surroundings of the clusters to $\geq 10$ virial radii ($r_{200c}$), making them also suitable to study the large scale environmental influence on galaxy evolution in and around clusters. To ensure that these clusters are centred at the peak of the local density field, there was an additional selection criteria of having no more massive halo within 30 pMpc or 20 $\times$ $r'_{200c}$ (whichever value is larger, $r'_{200c}$ is the virial radius of the neighbouring more massive halo). 

The AGNdT9 variant of the EAGLE simulation code \citep[see also \citealt{crain2015eagle}]{schaye2014eagle} was used to run the Hydrangea simulations. A significantly modified version of the smooth particle hydrodynamic (SPH) simulation code  \textsc{gadget-3}  \citep{springel2005cosmological}, this code uses an updated hydrodynamics scheme \citep{schaye2014eagle, schaller2015eagle} and a range of subgrid physics models to simulate astrophysical processes that originate below the resolution scale of the simulation. These astrophysical processes include star formation (based on the Kennicutt-Schmidt relation as in \citealt{schaye2008relation}, with the metallicity-dependent star formation threshold from \citealt{schaye2004star}), energy feedback from star formation in thermal, stochastic form to limit numerical cooling losses \citep{dalla2012simulating}, radiative cooling, photo-heating and re-ionization \citep{wiersma2009effect}, and metal enrichment from stellar evolution \citep{wiersma_2009_metallicity}. Models based on \citet{springel2005cosmological}, \citet{rosas2015impact}, and \citet{schaye2014eagle} are used to simulate the seeding, growth of, and feedback from supermassive black holes at galactic centres. The efficiency scaling of star formation feedback was calibrated against low-redshift observations of the field stellar mass function (SMF) and stellar mass--size relation \citep{crain2015eagle}.

 A flat $\Lambda$CDM cosmology is assumed in the Hydrangea suite (and EAGLE), with parameters taken from the \textit{Planck} 2013 results, combined with baryonic acoustic oscillations, polarization data from WMAP, and high multipole moment experiments \citep{planck2013}: Hubble parameter H$_0$ = 67.77 kms$^{-1}$Mpc$^{-1}$, dark energy density parameter $\Omega_{\Lambda} = 0.693$, matter density parameter $\Omega_\textrm{M} = 0.307$, and baryon density parameter $\Omega_\textrm{b} = 0.04825$. 

The primary output from each simulation are 30 snapshots spaced (mostly) equally between $0<z<14.0$, with a time step of $\Delta t$ = 500 Myr. In each of these snapshots, structures were identified through two consecutive steps with the \subfind{} code \citep[see also \citealt{springel2001}]{dolag2009}, which determines the gravitationally bound stellar, dark matter, and gas content of each identified object. In the first step, a friends-of-friends (FoF) algorithm with linking length equal to 0.2 times the mean inter-particle separation is used to identify spatially disjoint groups of DM particles. Baryon particles are attached to the FoF group (if any) of their nearest DM particle \citep{dolag2009} and FoF groups with fewer than 32 DM particles are discarded. In the second step, gravitationally bound candidate `subhaloes' within each FoF group are identified as locally over-dense regions. Particles that are not gravitationally bound to each of the subhalo candidate are then iteratively removed, and particles in the FoF group that are not part of any subhalo, but still gravitationally bound to the group, are considered as the `background' or `central' subhalo (see \citealt{bahe2017hydrangea} and \citealt{Bahe_et_al_2019} for more details). In this paper, we refer to all the subhaloes other than the central subhaloes as the `satellites'.  We use the terms `subhaloes' and `galaxies' interchangeably (i.e. both the terms refer to the baryonic and DM contents within the structure together). The most massive FoF group in each simulation is identified as the galaxy cluster whose properties we study in this work. Its (dominant) background subhalo is identified as the central, brightest cluster galaxy (BCG), all others are referred to as satellite galaxies. 

\begin{center}
	\begin{figure}
		\includegraphics[width=  \columnwidth]{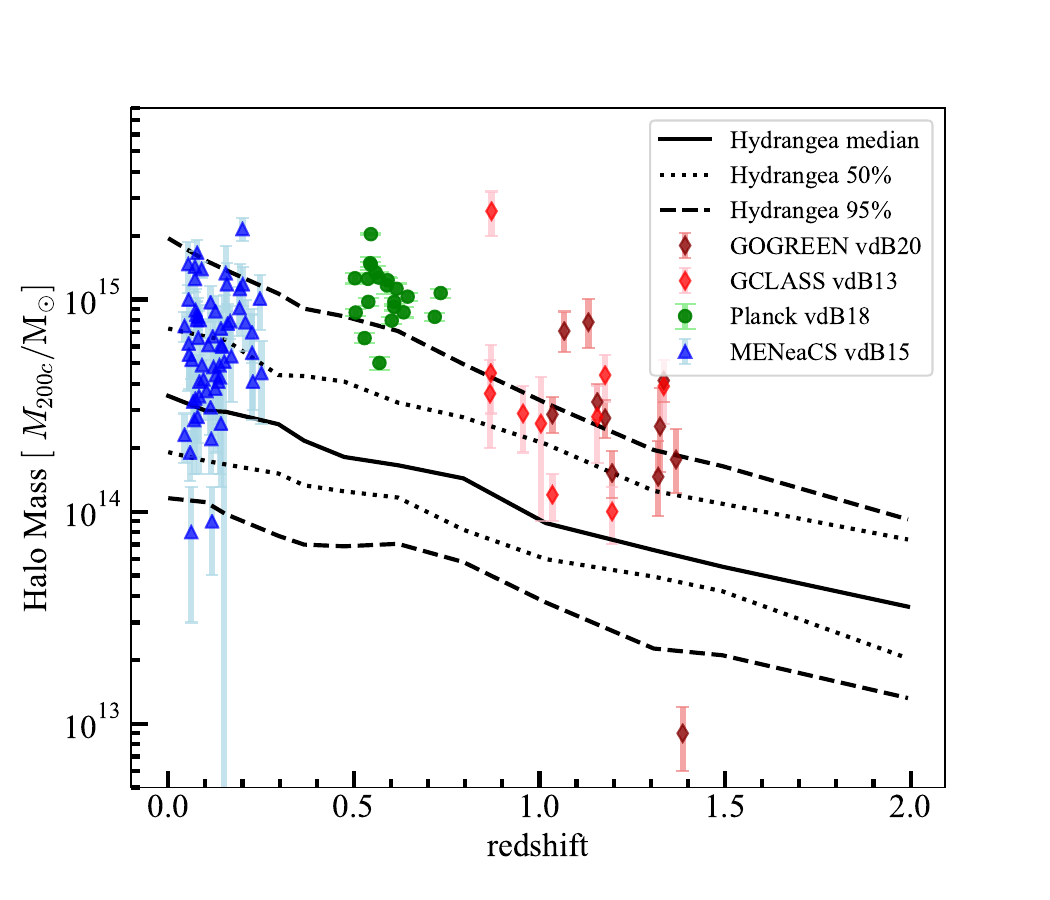}
		\caption{The time evolution of the cluster halo mass $M_{200\textrm{c}}$ (growth curves)  of the Hydrangea sample along with the observational data samples used in this paper. The solid black line shows the median cluster halo mass of the Hydrangea sample in the redshift range $0 \leq z \leq 2$. The dotted and dashed black lines include 50\% and 95\% of clusters from the Hydrangea sample, respectively. Brown diamonds represent the GOGREEN cluster sample \citep{Old_2020_GOGREEN}, red diamonds the GCLASS clusters \citep{van2013environmental, vdb14}, green squares the Planck-SZ sample \citep{van2018stellar}, and blue triangles the MENeaCS cluster samples from \citet{van2015evidence}. While the MENeaCS, GCLASS, and GOGREEN samples are mostly overlapping the Hydrangea clusters in mass at their respective redshifts, some outliers are significantly more massive, as is the majority of the Planck-SZ clusters. In Section \ref{smfschech}, we develop an extrapolation scheme to take these mass differences into account.}
		\label{massdist}
	\end{figure}
\end{center}

Figure~\ref{massdist} shows the mass evolution of the 24 simulated clusters over the redshift range $0 \leq z \leq 2$. To construct the growth curves shown by the black lines in Fig.~\ref{massdist}, we select at each redshift the most massive cluster halo from each of the 24 Hydrangea simulations. At the redshifts we consider ($z \leq 2$), these are mostly the actual main progenitors of the most massive $z = 0$ clusters: even at $z = 2$, only five of the most massive clusters in their simulation at this time are not going to remain the most massive one\footnote{In addition, two of them are subdominant progenitors of their $z = 0$ cluster.} until $z = 0$. Selecting the actual main progenitors instead would decrease the median halo mass by only 25 per cent at $z = 2$, and 6 per cent at $z = 1$; our results are therefore insensitive to this choice.


As expected \citep[e.g.][]{van2015evidence}, the distribution of Hydrangea cluster masses decreases steadily towards higher redshift, but maintains its roughly one order of magnitude spread. At $z = 1$ (2), the median mass is lower by a factor of $\approx$3 (10) than at $z = 0$. This broadly mimics the distribution of the observed cluster masses from our comparison samples (coloured symbols in Fig.~\ref{massdist}), as we discuss in Section \ref{obsdata}.

\subsection{Verifying the consistency between satellite stellar masses from \subfind{} and observations}
\label{sec:mstar_subfind}
Previous works have shown that during the subhalo identification step, \subfind{} can assign particles incorrectly to the central rather than to a satellite subhalo \citep[e.g.][]{muldrew2011subfind}. This can artificially suppress the mass of subhaloes near the centre of a galaxy cluster or, in extreme cases, even lead to them being missed completely. \citet{Bahe_et_al_2019} found that, in Hydrangea, only $\leq 5$ per cent of cluster galaxies with $M_\textrm{tot}^\textrm{peak}>10^{10}\textrm{M}_{\odot}$  are missed completely by \subfind{} (where $M_\textrm{tot}^\textrm{peak}$ is the peak total mass of the DM, stars, and black hole particles over the entire life of the galaxy). However, the masses assigned to identified satellite subhaloes may still be significantly underestimated \citep[e.g.][]{Behroozi_et_al_2013,Canas_et_al_2019}. Before studying the stellar mass in the Hydrangea clusters in detail, we therefore test whether the estimated stellar masses of the satellite subhaloes are comparable with what would be inferred from the observations.

\begin{figure}
	\includegraphics[width=  \columnwidth]{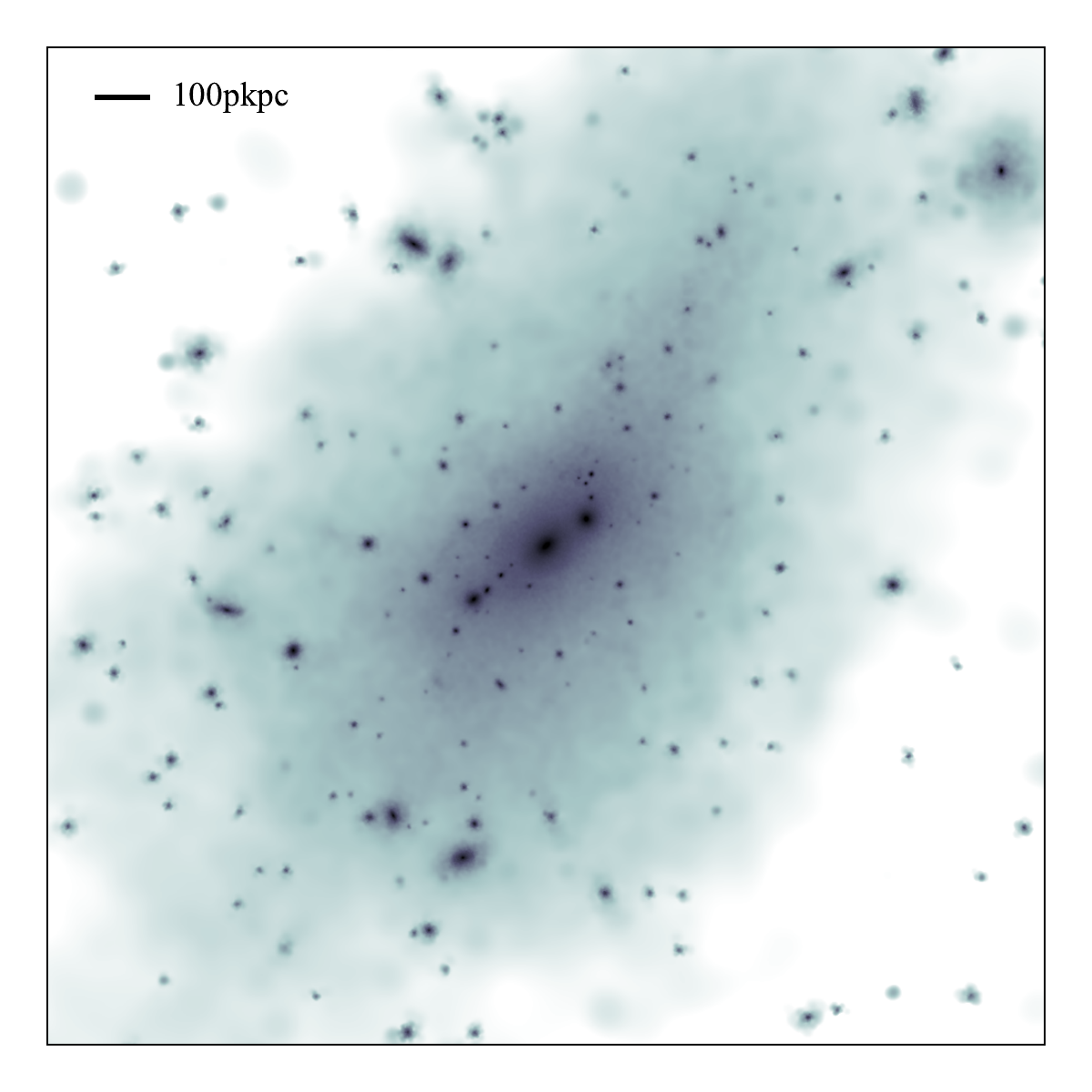}
	\caption{An example of a synthetic image of the surface mass density distribution of the stellar component of cluster CE-0 from the Hydrangea simulations. It is centered on the cluster potential minimum and extends to 1 pMpc on each side from the center along both axes. The image is obtained by projecting the star particles along the y-z axis of the simulation frame. There is no noise added in this figure, and the pixel values range from $\sim$$10$ to $\sim$$10^{9}\,\msun\,\mathrm{kpc}^{-2}$.}
	\label{synimg}
\end{figure}

For this purpose, we produce synthetic images from the simulated cluster snapshot data at $z=0$. We take the three dimensional (3D) stellar particle positions and calculate the 2D surface mass density distribution, projected along each principal axis of the simulation, for the central region of each of the 24 simulated clusters (see Fig.~\ref{synimg} for an example). Each image has a size of $1000\times1000$ pixels of side length 2 Mpc and is centred at the position of the potential minimum of the central cluster\footnote{The images therefore stretch to $\sim 0.5 \times r_{200\textrm{c}}$ from the cluster centre.}. The image value at each pixel, in units of M$_{\odot}$, represents the stellar mass surface density, which would be equivalent to the luminosity under the assumption of a fixed mass-to-light ratio.

As the output from the simulation is noise-free (except for Poisson noise from the finite number of particles), these raw images cannot be directly fit with observational techniques. We therefore add artificial noise at an RMS level of $1.5\times 10^6$ M$_{\odot}$ per pixel to mimic the depth of the cluster images from the Multi-Epoch Nearby Cluster Survey (MENeaCS, see Section \ref{menobs}) at $z<0.25$. The RMS noise level is determined by converting the noise level of light from the observations to stellar mass to be used for the simulated image. We then use Source Extractor (\textsc{SExtractor}, \citealt{bertin1995sourceex, 2010ascl.soft10064B}) to detect the objects (in our case, galaxies) in the image. The detection criterion is set such that at least 3 adjacent pixels have a flux density that is greater than $4\sigma$ in comparison to the local background. Here, the threshold value of $4\sigma$ corresponds to $\sim10^7$ M$_{\odot}$, which makes sure that we are not selecting spurious line-of-sight collections of star particles as galaxies, and that the threshold is small enough to not exclude any galaxies above a stellar mass of $10^9\,\msun$. The output flags for nearby bright neighbours and de-blending are taken into account to avoid overestimating the mass from the integrated flux of a faint object in case it is close to a very bright galaxy, especially in case of the smaller satellites near the BCG.  


Using the position of sources detected in this way, we matched them to the subhaloes identified by \subfind{}. We consider a \textsc{SExtractor} source and subhalo to be matched if their projected distance is less than 1.5 pixels ($=3$ pkpc). We take the 3D subhalo stellar mass within 30 pkpc from the subhalo center of potential for each of the matched subhaloes as the mass from \subfind{}. For field galaxies, this 3D aperture has previously been shown to mimick the 2D Petrosian aperture that is frequently used in observations \citep[][see their sec. 5.1.1. and fig.~6]{schaye2014eagle}. From the synthetic images, we estimate the mass from the integrated \texttt{FLUX\_AUTO} output for each object from \textsc{SExtractor}. Then, we compare the stellar mass function from the \subfind{} stellar mass and the mass retrieved by \textsc{SExtractor} from the images. To account for any projection bias that may occur from taking the 2D projected data in this test, we project each cluster separately along the $x$, $y$, and $z$ axes, and repeat the above procedure for each projection.

\begin{figure}	
	\includegraphics[width=\columnwidth]{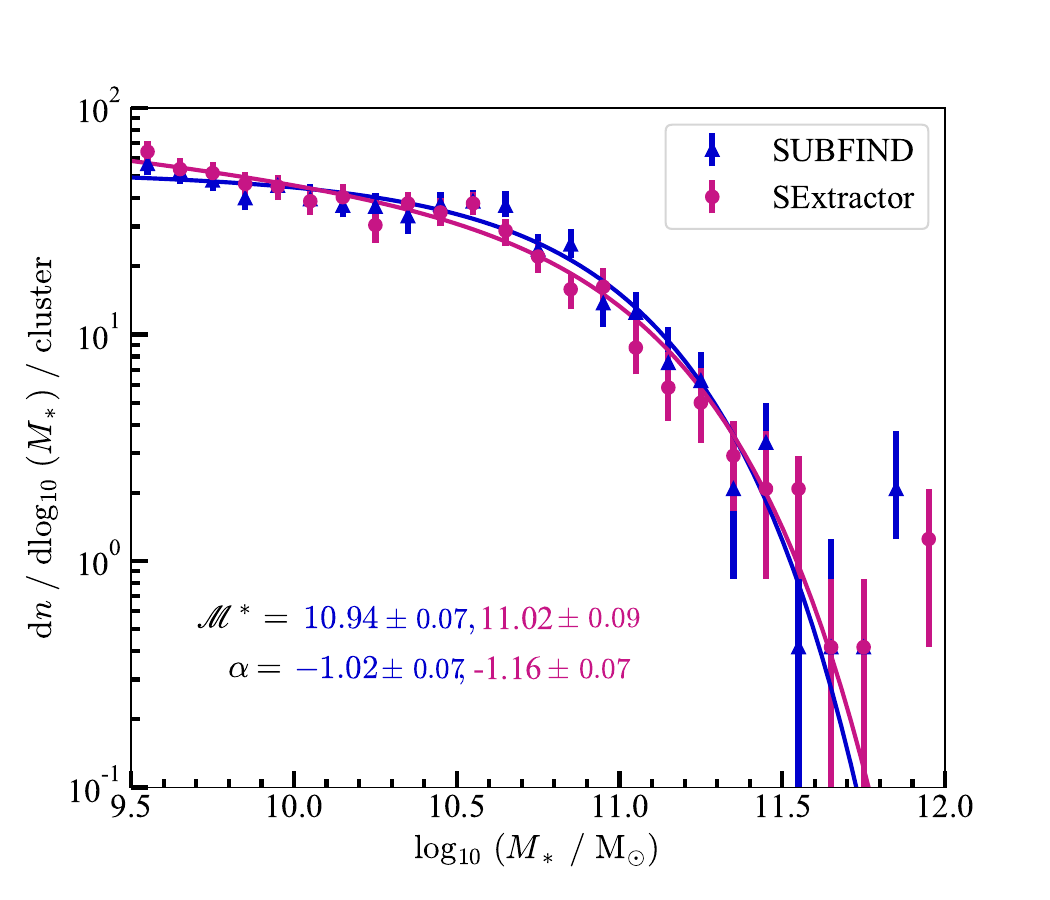}
	\caption{The galaxy stellar mass function (GSMF) of the simulated clusters (coloured symbols) and their best-fit Schechter functions (solid lines). The blue points are obtained from the subhalo stellar mass measured by \subfind{} within 30 pkpc from the centre of potential of each subhalo; magenta points are obtained from the estimated stellar mass of each galaxy by running \textsc{SExtractor} on synthetic images (see text for details). The error bars are obtained by bootstrap re-samplings of the stack of galaxies in each sample. Both the non-parametric mass functions and their Schechter fits agree within statistical uncertainties, indicating that the \subfind{} mass measurement is consistent with the observational approach. }
	\label{synismfmg}
\end{figure}

Figure~\ref{synismfmg} shows the galaxy SMFs obtained from the \subfind{} output (blue) and the synthetic images (magenta). The error bars here represent the Poisson errors obtained from 100 bootstrap re-samplings of the stack of galaxies in each sample. For all three projections, the stellar mass functions using galaxy masses from \subfind{} and those from the synthetic images agree within their uncertainties. The best-fit Schechter functions \citep{schechter1976analytic} for the SMFs (solid lines, see Section \ref{smfschech} for details) in Fig.~\ref{synismfmg} indicate that the \subfind{} and \textsc{SExtractor} outputs agree well within the error bars. To test for different depths from observations at different redshifts, we repeated the procedure for RMS noise levels of $7.5\times 10^5$, $3.0\times 10^6$, and $6.0\times 10^6\,\msun\,\mathrm{pixel}^{-1}$. {\modtext The SMFs resulting from these SExtractor outputs agree similarly well with the \subfind{} SMF as the one shown in Fig.~\ref{synismfmg}, albeit with a mild dependence of the best-fit SMF parameters on the noise level (see Appendix \ref{app:noise_lev_test})}. We therefore conclude that the subhalo stellar mass within 30 pkpc as measured by the \subfind{} code can be reliably used for the following analysis.


\subsection{Total stellar mass fractions predicted at different redshifts}
\label{sm_to_hm}

As mentioned above, the Hydrangea simulations successfully reproduce the observed SMF of galaxy clusters in the local Universe, at $z\approx0.1$ \citep{bahe2017hydrangea}. As a first test, we explore how well the Hydrangea suite reproduces the observed total stellar mass content at higher redshifts, up to $z = 1.4$. In recent studies, the stellar content of galaxy clusters (and also their total baryon content) has been observed to depend strongly on cluster halo mass, but at best weakly with redshift up to $z\approx 1.3$ \citep{chiu2016,chiu2016b,chiu2018}. \citet{chiu2018} explored the mass scaling relations of the stellar content, as well as the intracluster medium, and total baryon mass in their sample of 91 Sunayev-Zel'dovich effect (SZE) selected galaxy clusters from the South Pole Telescope (SPT-SZ) survey in the redshift range $0.2<z<1.25$. They compared their scaling relations with other observational data, and showed that despite the residual systematic uncertainties among different datasets, the qualitative trend of the scaling relations does not vary significantly (see e.g. their fig. 6). As a first test, we compare the Hydrangea clusters with the scaling relation presented in \citet{chiu2018} to investigate whether the simulations can reproduce these fundamental and robustly measured observational features. 
 
\begin{figure}
	\includegraphics[width=  \columnwidth]{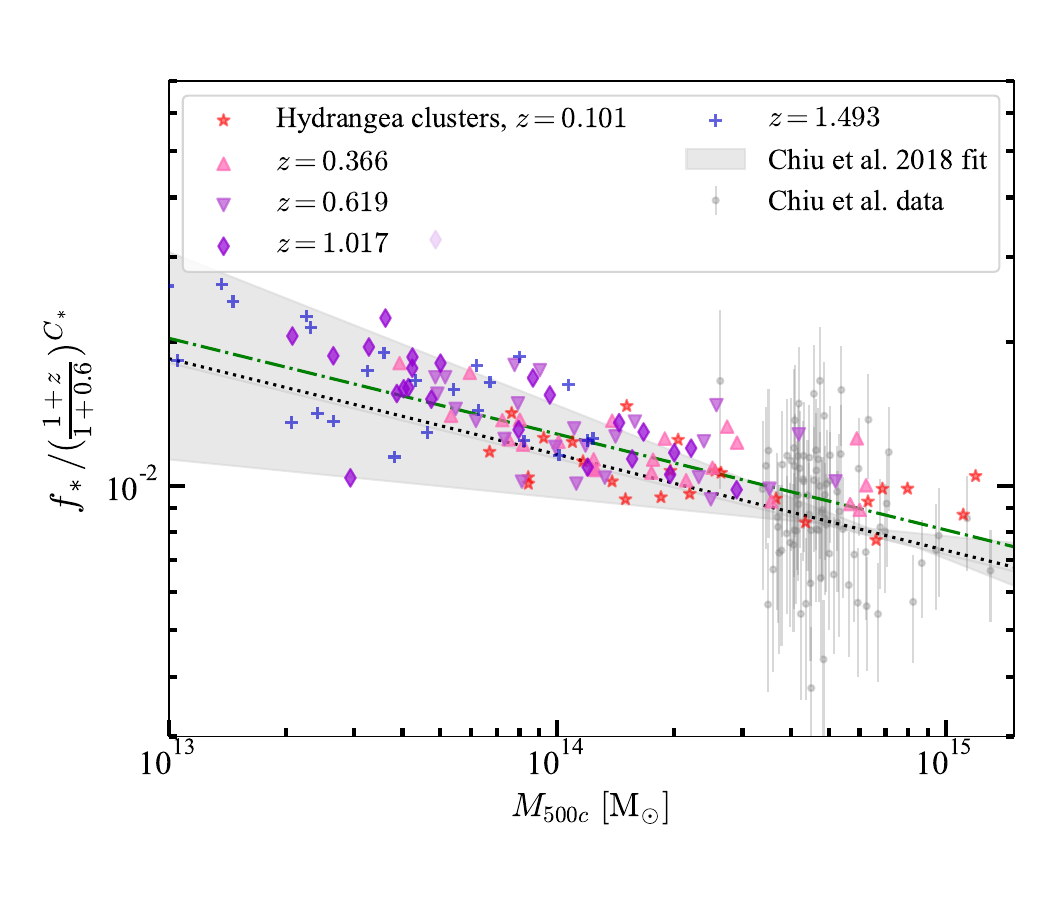}
	\caption{The scaling relation of the total stellar mass fraction with respect to the cluster halo mass. Different colored symbols (except for the grey ones) represent individual Hydrangea clusters at five different redshifts, as specified in the legend. The green dash-dotted line shows the best-fit scaling relation to all these points (see text for details). {\modtext The grey data points and corresponding error bars are the SPT clusters from \citet{chiu2018}. The dotted black line and grey shaded region give the corresponding best-fit relation from \citet{chiu2018} and its 1$\sigma$ confidence interval, respectively. } The simulations and observations agree well within uncertainties, confirming that the Hydrangea clusters contain realistic amounts of stellar mass over the majority of cosmic history.}
	\label{fig:chiuftest}
\end{figure}

Figure~\ref{fig:chiuftest} shows the scaling relation of the total stellar mass fraction within the (projected) radius $R_{500c}$ of the Hydrangea clusters at redshifts $0.1 < z < 1.5$ with respect to their total halo mass ($M_{500c}$, to match what is used by \citealt{chiu2018}), along with the \citet{chiu2018} scaling relation within its $1\sigma$ confidence interval. To compare with these results, we compute the stellar mass fraction in the simulations from the summed stellar masses of all galaxies (as computed by \subfind{}, within a 30 pkpc radial aperture) with $M_\mathrm{star} > 10^{10}\,\msun$ that lie within a 2D projected radius of $R_{500c}$ from the centre of potential of each cluster.


As in \citet{chiu2018}, we fit the stellar mass fractions with a relation of the form
\begin{equation}
\label{eq:scaleeqn}
\frac{f_\star}{{\left(\frac{1+z}{1+z_{\textrm{piv}}}\right)}^{C_\star}} = A \times \frac{10^{12}\,\textrm{M}_{\odot}}{M_{500c}}  \times \left({\frac{M_{500c}}{M_{\textrm{piv}}}}\right)^B, 
\end{equation}

\noindent where $f_\star$ is the stellar mass fraction, $z_{\textrm{piv}}$ the pivot redshift, and $M_{\textrm{piv}}$ the pivot halo mass, and $A$, $B$, and $C_\star$ are free parameters; we follow \citet{chiu2018} and use $z_{\textrm{piv}} = 0.6$ and $M_{\textrm{piv}} = 4.8\times10^{14}\,\msun$, respectively\footnote{Following \citet{chiu2018}, we divide $f_\star$ by the factor $\left( (1+z) / (1+z_\mathrm{piv}) \right)^{C_\star}$ to account for the (slight) redshift dependence of the stellar mass fractions.}. We fit the Hydrangea data with Eqn.~\ref{eq:scaleeqn} for parameters $A$ and $B$ by maximizing the log likelihood value. The redshift scaling index $C_\star$ is kept fixed at 0.05, the best-fit value of \citet{chiu2018}.

The best-fit parameters, represented by the green dash-dotted line in Fig.~\ref{fig:chiuftest}, and their $1\sigma$ confidence intervals, are $A = 4.5\pm0.8$, and $B = 0.79\pm0.1$. In comparison, the observational best-fit parameters \citep{chiu2018} are $A = 4.0\pm0.28$, and $B = 0.8\pm0.12$, respectively, as indicated by the dotted black line and gray band in Fig.~\ref{fig:chiuftest}. The deviation, $\Delta A = 0.5$ and $\Delta B = 0.01$, is within the $1\sigma$ uncertainty in both cases. The slightly higher-than-observed normalization for the Hydrangea clusters is in fact not too surprising, given that \citet{bahe2017hydrangea} showed that the simulated BCGs at $z = 0$ are a factor of $\sim$3 too massive. We also note that the cluster-to-cluster scatter in Fig.~\ref{fig:chiuftest} is similar to what is shown in fig.~6 of \citet{chiu2018}, where they show their cluster sample along with other observed cluster samples at a range of redshifts.

From the comparison presented in Fig.~\ref{fig:chiuftest}, we therefore conclude that the Hydrangea simulations predict a total stellar content within galaxy clusters that is quantitatively consistent with observations out to $z\approx 1.5$. Based on this fundamental success, we explore in the subsequent sections whether the Hydrangea suite can reproduce the distribution of the total stellar mass in individual galaxies, i.e., the stellar mass function (SMF) at redshifts in the range $0 < z < 1.4$.

\section{Observational Data}
\label{obsdata}
To test the accuracy of the GSMF and the radial distribution of stellar mass predicted by Hydrangea, we compare the simulations with several recent observational data sets at a range of redshifts. These include the MENeaCS and CCCP cluster samples at $z\approx 0.15$ as analysed by \citet{vdb14} and \citet{van2015evidence}, the Planck-SZ clusters at $z\approx 0.6$ \citep{van2018stellar}, GCLASS clusters at $z \approx 1.0$ as presented by \citet{van2013environmental}, and from the GOGREEN survey at $z\approx1.3$ \citep{vanderburg2020GOGREEN}. In this section, we summarize the aspects of these different data sets and how their relevant characteristics are reproduced in our simulation analysis. To illustrate their halo mass range and its relation to Hydrangea at the respective redshifts, all observed clusters used for comparison in Sections \ref{smfdis} and \ref{denprofsec} are shown as symbols in Fig.~\ref{massdist}; we comment on the mass distribution of each sample in the respective sub-sections below.

\subsection{MENeaCS and CCCP}
\label{menobs}
The sample from the the Multi-Epoch Nearby Cluster Survey (MENeaCS) and the Canadian Cluster Comparison Project (CCCP) consists of 60 massive clusters in the local Universe \citep[$z<0.25$,][]{van2015evidence}. For this sample, deep \textit{ugri}-band photometry (with a median limiting \textit{g}-band magnitude of 24.8) is used together with spectroscopic observations to determine cluster membership and to estimate the dynamical mass of the clusters. The cluster halo masses, defined as $M_{200\textrm{c}}$, range from $8\times 10^{13}\,\msun$ to $2.15 \times 10^{15}\,\msun$, with a mean of $6.8\times 10^{14}\,\msun$. In comparison, the average halo mass of the Hydrangea clusters at similar redshift is $\approx$5$\times 10^{14}\,\msun$; as can be seen from Fig.~\ref{massdist} (blue triangles), the range of simulated clusters masses overlaps well with these observed samples despite the slightly different averages.

From these clusters, we use the radial stellar mass density distribution of their satellite galaxies presented in \citet{van2015evidence}. These authors construct an ensemble cluster by stacking the satellites around individual clusters, with their (2D) radial distances normalised by their respective $R_{200c}$, and determine the 2D density profile in this stack over the (projected) radial range from 0.1 to $2 \times R_\mathrm{200c}$. We apply the same stacking procedure and radial range in our simulation analysis.

\subsection{\textit{Planck}-SZ}
\label{plszobs}
The observed clusters at $z\approx0.6$ from \citet{van2018stellar} were primarily detected with \textit{Planck} via the Sunayev-Zeldovich (SZ) effect and later confirmed with follow-up observations. The Planck-SZ selected clusters are part of the most massive cluster samples from that epoch. \citet{van2018stellar} present the GSMF of 21 \textit{Planck}--SZ selected galaxy clusters. The cluster halo masses are obtained by applying the $M - Y_\textrm{X}$ relation to deep X-ray maps, and range between $M_\mathrm{500c} = 3.17 \times 10^{14}\,\msun$ and $1.28 \times 10^{15}\,\msun$ with a mean of $6.9\times 10^{14}\,\msun$. This is significantly higher than the average of the Hydrangea clusters at similar redshift ($1.6 \times 10^{14}\,\msun$); as Fig.~\ref{massdist} shows, all but two of their clusters (green circles) lie above the 95 per cent halo mass interval of the simulations. We overcome this mismatch through a mass extrapolation scheme that we apply to the simulations, as described in Section \ref{smfschech}. 

We use the GSMF of the Planck-SZ clusters as published by \citet{van2018stellar} for comparison with our simulations. The brightest cluster galaxy (BCG), selected in the K$_s$ band within a $1'$ location limit from the X-ray peak for each cluster, is excluded and the GSMF is constructed by stacking all other galaxies located within $2 \times R_\mathrm{500c}$, with a stellar mass limit of $10^{9.5}\,\msun$ for the total galaxy population. For statistical background subtraction and comparison with the field environment, they use data from the COSMOS/Ultra VISTA field at similar redshift range \citep{muzzin2013public}. 

\subsection{GCLASS and GOGREEN}
\label{gclobs}
The GCLASS cluster sample reported in \citet{van2013environmental} consists of 10 red-sequence selected rich galaxy clusters at $0.86 < z < 1.34$ in the Gemini Cluster Astrophysics Spectroscopic Survey (GCLASS). The cluster halo masses, defined as $M_\mathrm{200c}$, are estimated from the line-of-sight velocity dispersion of the spectroscopic targets and range from $1.0 \times 10^{14}\,\msun$ to $2.61 \times 10^{15}\,\msun$ with a mean of $5.3 \times 10^{14}\,\msun$. As with the \textit{Planck}-SZ clusters, this is a factor of $\approx$5 higher than the equivalent Hydrangea average ($9.4 \times 10^{13}\,\msun$). However, the wide range in GCLASS halo masses means that more than half of the sample nevertheless overlaps with the Hydrangea clusters in mass (see the red diamonds in Fig.~\ref{massdist}). We apply the same extrapolation scheme as for the Planck-SZ clusters to account for this mass mismatch in our comparison of the GSMF. \citet{van2013environmental} construct the GSMF by stacking the satellite galaxies located within a projected radius of 1 pMpc from the cluster centre with a lower stellar mass limit of $10^{10}\,$M$_{\odot}$. 

At a similar redshift range, ($1.0 < z < 1.4$) \citet{vanderburg2020GOGREEN} present the GSMF of 11 galaxy clusters from the Gemini Observations of Galaxies in Rich Early Environments (GOGREEN) survey. The cluster halo masses range from $M_\mathrm{200c} = 1.0 \times 10^{13}$ to $7.8 \times 10^{14}\,\msun$, with a mean of $3.2\times 10^{14}\,\msun$. This is similar to the GCLASS sample (also in terms of cluster-to-cluster scatter, see the brown diamonds in Fig.~\ref{massdist}), so that again only a small extrapolation correction is required in our comparison below. \citet{vanderburg2020GOGREEN} construct the GSMF by stacking all the cluster galaxies (including the BCGs) with $M_\star \geq 10^{9.75}\,\msun$ that are located within a projected radius of 1 pMpc from the cluster centre.

\section{Galaxy Stellar Mass Function}
\label{smfdis}

As we have shown in Section \ref{sm_to_hm}, the Hydrangea simulations predict a total stellar mass fraction in clusters that is consistent with observations out to $z \approx 1.5$. We now perform the more stringent test of how realistically this mass is distributed over satellites of different masses, i.e. the accuracy of the predicted satellite GSMF. We first describe our approach for dealing with the (moderate) offsets between the masses of simulated and observed clusters as discussed above (Section \ref{smfschech}), and then confront simulations and observations in Section \ref{sec:testingmodelwdata}.

\subsection{Accounting for differences in cluster halo mass}
\label{smfschech}
As shown in Fig.~\ref{massdist}, the mass range of the Hydrangea clusters is not exactly matched to the observational comparison samples, with many of the more massive observed clusters not having any similarly massive analogue in the simulations. Although the offsets are not huge (factors of $\lesssim 5$, see Section \ref{obsdata}), the tight correlation between cluster richness and mass \citep[e.g.][]{Yee_Ellingson_2003, Budzynski2012MNRAS, Pearson_et_al_2015} means that they nevertheless have to be accounted for to enable a meaningful test of the GSMF. We do this by means of an extrapolation model based on the \citet{schechter1976analytic} fitting formula:
\begin{equation}
\Phi(M) = \ln(10) \, \Phi^* \, (M/\mathscr{M}^*)^{(1+\alpha)}\, \mathrm{e}^{-M/\mathscr{M}^*}, 
\label{eq:schechter}
\end{equation}
with normalization $\Phi^*$, characteristic mass $\mathscr{M}^*$, and $\alpha$ setting the (logarithmic) slope at the low-mass end.

\begin{figure}
	\includegraphics[width=\columnwidth]{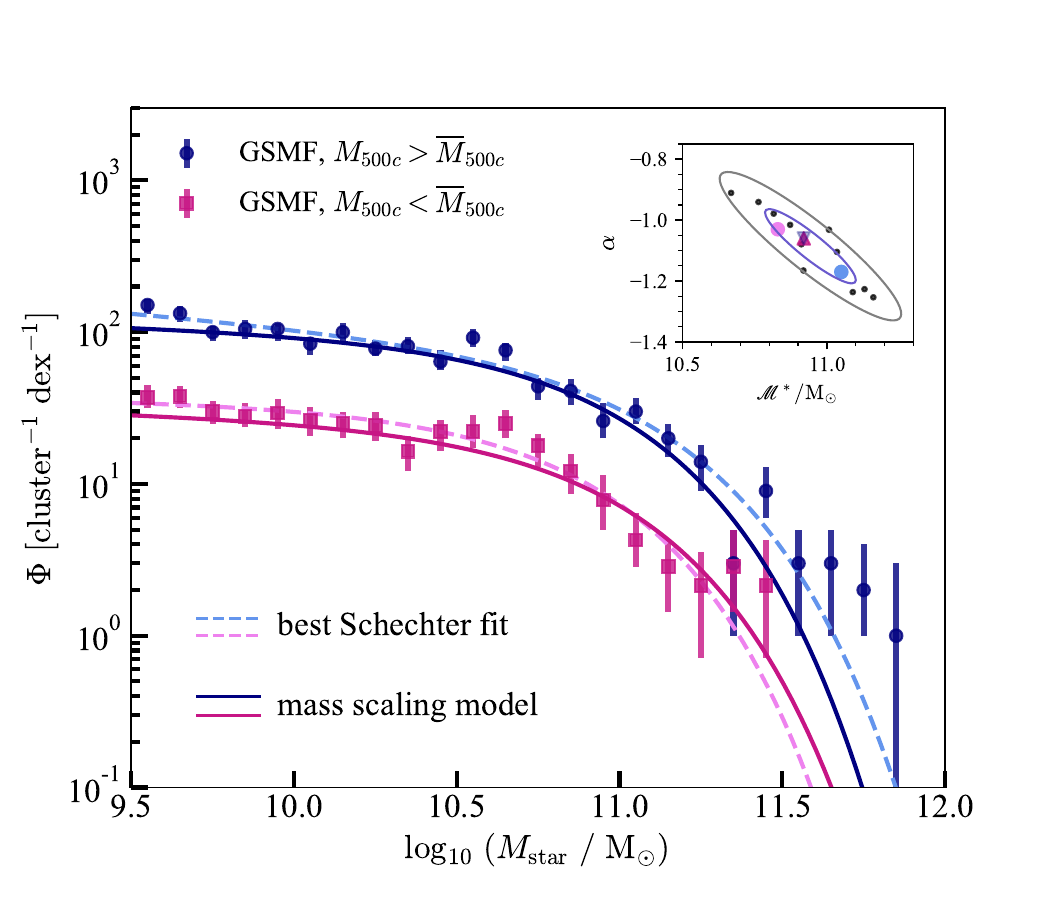}
	\caption{The stellar mass function (SMF) of the Hydrangea clusters in 2 mass bins at $z\approx0.6$. The dark blue circles (magenta squares) show the stacked SMF of clusters with $M_{500c}$ above (below) the mean $\overline{M}_\mathrm{500c} = 1.6 \times 10^{14}\, \msun$ of all the clusters; error bars represent $1\sigma$ uncertainties from bootstrapping (see text for details). Dashed lines show the best-fit Schechter functions for each set of points, while solid lines represent the output from our mass scaling model (details in Section \ref{smfschech}) at the mean $M_\mathrm{500c}$ of each stack. The inset in the upper-right corner shows the correlation between the Schechter parameters $\alpha$ and $\mathscr{M}^*$ for the two stacks (light blue/magenta circles), the mass scaling model (dark magenta upward/dark blue downward triangles), and the 11 individual fits described in Section \ref{smfschech} (black circles). The blue and gray ellipses in the inset represent the $1\sigma$ and $2\sigma$ uncertainties for the fitted parameters. The simulated SMF is well-described by the mass-scaled Schechter function, but with some degeneracy between its parameters $\alpha$ and $\mathscr{M}^*$.}
	\label{fig:smf2mass}
\end{figure}

In Fig.~\ref{synismfmg}, we had already seen that the $z = 0$ GSMF predicted by Hydrangea is well-fit with a Schechter function. Similarly good fits are achieved at higher redshift: as an example, we show in Fig.~\ref{fig:smf2mass} the predicted GSMF at $z = 0.6$, splitting our cluster sample into a low- and high-mass subset at the mean $\overline{M}_\mathrm{500c} = 1.6 \times 10^{14}\, \msun$. The satellite GSMF for the low (high) mass clusters is shown with magenta squares (blue circles). Error bars indicate $1\sigma$ confidence intervals from 100 bootstrap re-samplings; for each of these, we draw from all galaxies with $M_\star \geq 10^{9.5}\,\msun$ in the cluster ensemble a Poisson-distributed number with mean equal to the total number of galaxies in the ensemble (with replacement). We use the Monte-Carlo Markov Chain (MCMC) package \textsc{emcee} \citep{Foreman_Mackey_2013} to find the best-fit Schechter function parameters ($\Phi^*$, $\mathscr{M}^*$, $\alpha$) and their uncertainties.


These best-fit Schechter functions are plotted in Fig.~\ref{fig:smf2mass} as dashed lines in the same colour as the corresponding binned GSMF (see Table \ref{tab:fitted_params} for the corresponding values). Two key features are evident: the GSMF of the lower-mass clusters (magenta) is indeed significantly lower than for the more massive cluster set (blue), and each is well-described by its best-fit Schechter function (with the possible exception of the very high mass end, where the best fit for both sets falls slightly below their respective binned data points). As expected, the GSMF normalisations $\Phi^*$ of the high- and low-mass cluster stacks differ significantly (by a factor greater than 2), but the characteristic mass $\mathscr{M}^*$ and low-mass slope parameter $\alpha$ are similar in both stacks (to within 39 and 13 percent, respectively). Similarly good Schechter fits are obtained for the cluster ensembles at other redshifts (not shown).

Having established that the Schechter function is indeed a good representation for the satellite GSMF in Hydrangea, we then proceed to model its mass dependence for comparison to the observations. First, we compute and fit satellite mass functions over a range of cluster masses at each of the three required redshifts ($z = 0.619$, 1.017, and 1.308), in analogy what is shown in Fig.~\ref{fig:smf2mass}. At the high-mass end, we use the seven most massive clusters at each redshift individually, which yield good Schechter fits (reduced $\chi^2 \leq 2.0$ in most cases). For less massive clusters, however, we have found that their individual satellite counts are not high enough to derive robust Schechter fits. We therefore stack the lower mass clusters in sets of four at successively higher mass, and derive (and fit) composite mass functions from these stacks.

\begin{figure*}
	\includegraphics[width=2.1\columnwidth]{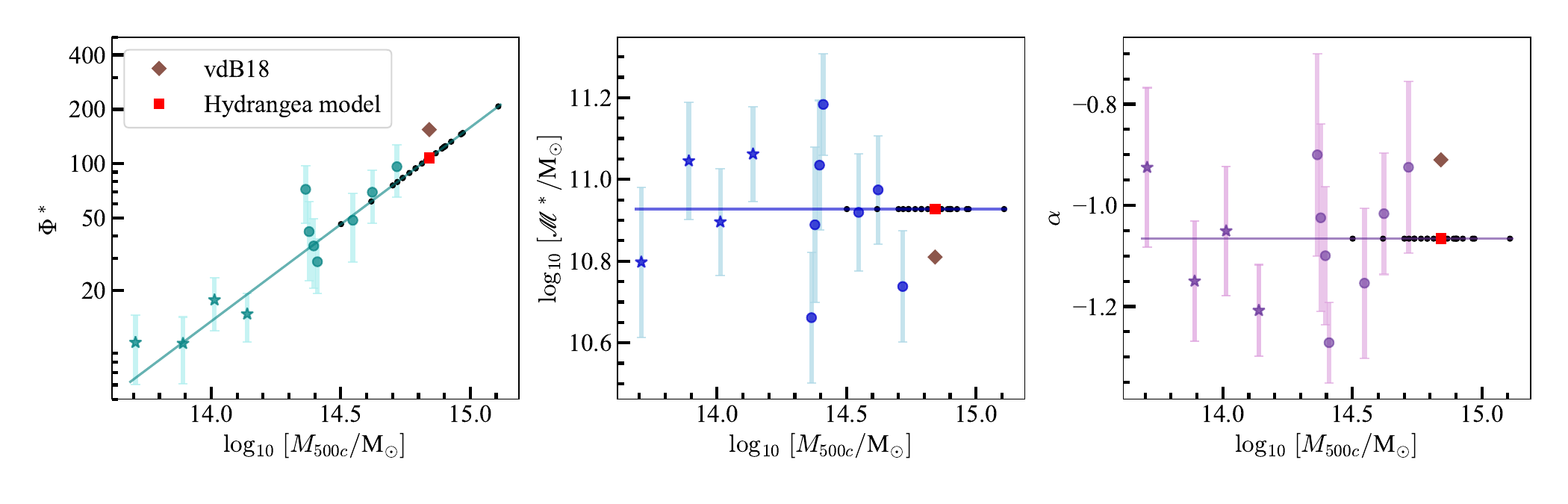}
	\caption{The best-fit Schechter function parameters of individual $z = 0.6$ simulated clusters (coloured circles), or stacks of four clusters with similar mass (coloured stars), as a function of cluster mass $M_\textrm{500c}$. From left to right, the panels show the variation in normalisation ($\Phi^*$), characteristic mass ($\mathscr{M}^*$), and low-mass slope (${\alpha}$), respectively. The best-fit power law model to the $\Phi^*$--$M_\mathrm{500c}$ relation, and the best-fit constant $\mathscr{M}^*$ and $\alpha$, are shown with solid lines. Black dots represent the corresponding parameters predicted by these fits at the mass of observed clusters from \citet{van2018stellar}; the red square gives their weighted average (see text). The actual ensemble SMF parameters from the observations are shown as brown diamonds. These agree with the simulation predictions within the cluster-to-cluster scatter.}
	\label{fig:schechterparams}
\end{figure*}

The result for $z \approx 0.6$ is shown in the three panels of Fig.~\ref{fig:schechterparams}, where the turquoise, blue, and purple symbols indicate the best-fit Schechter parameters $\Phi^*$, $\mathscr{M}^*$, and $\alpha$, respectively, as a function of cluster mass $M_\mathrm{500c}$. For the seven massive clusters that are fit individually, we show the best-fit parameters as circles, whereas stars represent the best-fit parameters for the ensemble stacks of lower-mass clusters, plotted at their mean $M_\mathrm{500c}$.


The left-hand panel shows the expected tight correlation between $\Phi^*$ and $M_\mathrm{500c}$. We fit it with a power-law in mass,
\begin{equation}
\label{eq:modelparam}
\Phi^* = a \cdot \left(\frac{M_\mathrm{cluster}}{M_{\textrm{piv}}}\right)^b,
\end{equation}
where $M_\mathrm{cluster}$ denotes the cluster mass (here $M_\mathrm{500c}$, motivated by the use of that value by \citealt{van2018stellar}), while $a$ and $b$ are the fitted parameters. The pivot mass $M_\mathrm{piv}$ is fixed to the mean $M_\mathrm{cluster}$ of all the objects included in the fit. The best-fit power law is shown as a turquoise solid line in Fig.~\ref{fig:schechterparams}; its index $b$ is 1.07, i.e. a close-to-linear relation between $\Phi^*$ and $M_\mathrm{500c}$. We note that the parameter $\Phi^*$ is closely related to the cluster richness; observationally, the cluster mass--richness relation follows a power-law with index $b$ in the range 0.61--1.3 at $z \approx 0.6$ \citep{Yee_Ellingson_2003,Budzynski2012MNRAS,Andreon2014,hurier2019evolution}. Encouragingly, the value we find from the simulations ($b = 1.07$) is well within this range.

The other two Schechter parameters, $\mathscr{M}^*$ and $\alpha$, show no clear dependence on $M_\mathrm{500c}$ (middle and right-hand panels of Fig.~\ref{fig:schechterparams}). As shown in the top-right inset of Fig.~\ref{fig:smf2mass}, there is however a noticeable correlation (i.e. degeneracy) between these two parameters: higher $\mathscr{M}^*$ correlates with lower $\alpha$. Since the majority of the individual best-fit parameters are consistent with a constant value of $\mathscr{M}^* = 10^{10.92}\, \msun$ and $\alpha = -1.06$, respectively, we therefore keep these values fixed in our mass dependence model.

In Fig.~\ref{fig:smf2mass}, we show the prediction from this mass-scaling model at the mean mass of the two cluster sets as solid lines. Although they do not trace the actual best-fit Schechter function (dashed lines) exactly -- especially at the low-$M_\star$ end, they are biased low by $\approx$10--20 per cent -- they clearly provide a comparably good description of the actual (binned) stellar mass functions.

We have repeated this procedure for $z = 1.0$ and $z = 1.3$, to enable GSMF comparisons to the \textit{Planck}-SZ, GCLASS, and GOGREEN observations. In Table \ref{tab:model_params}, we list the best-fit parameters $a$ and $b$ and the corresponding pivot masses $M_\mathrm{piv}$ for the power-law scaling of $\Phi^* (M_\mathrm{cluster})$, as well as the best-fit constant Schechter parameters $\mathscr{M}^*$ and $\alpha$, for each of these three redshifts. We note that, for compatibility with the respective observational data, the $z = 0.6$ parameters use $M_\mathrm{500c}$ as cluster mass, whereas those for $z = 1.0$ and $z = 1.3$ are based on $M_\mathrm{200c}$. 

The characteristic mass $\mathscr{M}^*$ and the low-mass slope $\alpha$ show only a slight variation between these three redshifts. The low-mass slope $\alpha$ increases by $\approx$14 per cent from $z = 0.6$ to $z = 1.3$, while $\mathscr{M}^*$ decreases by a similar amount, as expected from their anti-correlation at $z = 0.6$ (see the inset in Fig.~\ref{fig:smf2mass}). If real, the slight decrease in $\mathscr{M}^*$ could be interpreted as the most massive galaxies in the clusters ($M_\star >10^{11}\,\msun$) still growing at higher redshift. The power-law index for the scaling of $\Phi^*$ with cluster mass remains close to one (i.e. linear), although the pivot mass $M_\mathrm{piv}$ is (unsurprisingly) lower at higher redshift. 

\begin{table}
	\centering
	\caption{Hydrangea model parameters for the Schechter function parameters of the simulated clusters at $z\approx0.6, z\approx1.0$, and $z\approx1.3$.}
	\label{tab:model_params}
	\begin{tabular}{lccccr} 
		\hline
		{Parameters $\rightarrow$} & $M_{\textrm{piv}}$& \multicolumn{2}{c|}{$\Phi^*$} & $\log_{10}\mathscr{M}^*$ & $\alpha$ \\
		\cline{1-4}
        Redshift ($z$) $\downarrow$ & $[10^{14}~\textrm{M}_{\odot}]$ & $a$ & $b$ & $[\textrm{M}_{\odot}]$\\
		\hline
		$0.6$ & 1.59 &$22.10$ & $1.07$ & $10.92$ & $-1.06$\\
		$1.0$& 0.94 & $16.40$ & $0.77$ & $10.91$ & $-0.99$\\
		$1.3$ & 0.89 & $13.91$ & $1.05$ & $10.83$ & $-0.91$ \\
		\hline
	\end{tabular}
\end{table}

\subsection{Confronting simulations and observations}
\label{sec:testingmodelwdata}
Having constructed a model to scale the GSMF from Hydrangea to different cluster masses (Eqn.~\ref{eq:modelparam}, Tab.~\ref{tab:model_params}), we now compare the simulation predictions to observations. To begin, we use the data of \textit{Planck}-SZ selected clusters at $z\approx 0.6$ presented in \citet{van2018stellar}. We calculate the Schechter parameters predicted by our scaling model for each of the observed clusters, based on their halo mass, and show these as small black circles\footnote{By construction, these all lie exactly along the best-fit line for each parameter.} in each of the three panels of Fig.~\ref{fig:schechterparams}.

The ensemble average of each of these parameters is shown by the red squares in each panel. For $\mathscr{M}^*$ and $\alpha$ this is trivial, but for the normalization $\Phi^*$, care must be taken to weight the individual clusters in a similar way as in the observations. Since more massive clusters will contain more galaxies and hence contribute more to the ensemble average, we weight the individual predictions (black circles) by the normalization $\Phi^*$ before estimating the ensemble average. The corresponding best-fit parameters for the observed clusters from \citet{van2018stellar} are shown as brown diamonds. In all three cases, they lie close to the Hydrangea prediction, with an offset that is within the cluster-to-cluster scatter in the simulations. 

\begin{figure*}
	\includegraphics[width=2\columnwidth]{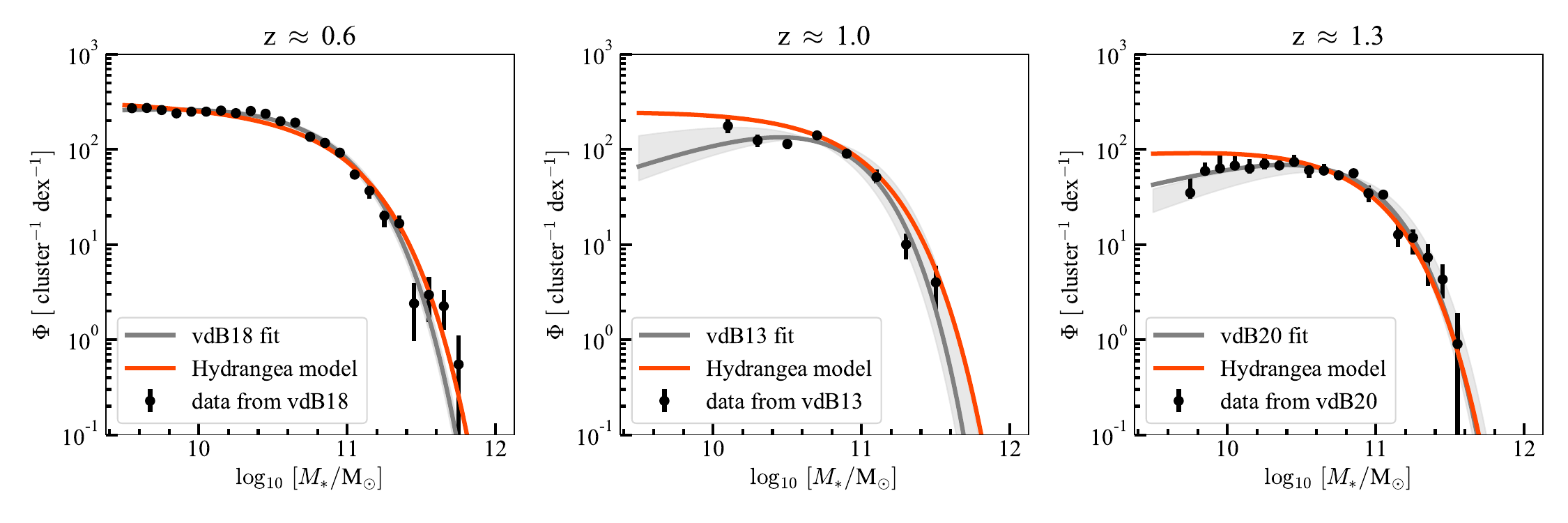}
	\caption{The stellar mass functions of the \textit{Planck}-SZ clusters at $z \approx 0.6$ \citep[][left]{van2018stellar}, the GCLASS clusters at $z \approx 1.0$ \citep[][middle]{van2013environmental}, and the GOGREEN clusters at $z \approx 1.3$ \citep[][right]{vanderburg2020GOGREEN} are shown as black circles with $1\sigma$ error bars, along with the best-fit Schechter functions from these papers (grey line, with $1\sigma$ uncertainty given by the light grey shaded region). Red lines represent the Schechter functions predicted from the Hydrangea simulations at the respective redshift and mean cluster mass. The simulations mostly agree with the observations, but predict too many low-mass galaxies ($M_\star \lesssim 10^{10.5}\,\msun$) at $z \geq 1$, by up to a factor of $\approx$2.}
	\label{fig:modelSMFz06}
\end{figure*}

In Fig.~\ref{fig:modelSMFz06}, we present our main result of the GSMF test by directly comparing the Schechter function predicted from the simulations (through our mass scaling model) to the observations. In the left-hand, central, and right-hand panels, respectively, we show as a red line the simulation predictions at $z = 0.6$ (with parameters as shown by the red squares in Fig.~\ref{fig:schechterparams}), $z = 1.0$, and $z = 1.3$. This is compared to the observationally measured GSMFs (black circles) from \textit{Planck}-SZ \citep{van2018stellar}, GCLASS \citep{van2013environmental}, and GOGREEN \citep{vanderburg2020GOGREEN} (black circles) and their best-fit Schechter functions (grey line), respectively.

For $z = 0.6$ (left-hand panel of Fig.~\ref{fig:modelSMFz06}), the predicted Schechter function agrees extremely well with its observational counterpart over the full stellar mass range we probe. At the highest masses ($M_\star>10^{11.2}\,\msun$), the Hydrangea prediction is marginally higher than the observational best-fit Schechter function from \citet{van2018stellar}. However, it is still fully consistent with the actual observational data points (black circles) within their ($1\sigma$) error bars.


The analogous comparison to GCLASS at $z \approx 1.0$ \citep[][middle panel of Fig.~\ref{fig:modelSMFz06}]{van2013environmental} and GOGREEN at $z \approx 1.3$ \citep[][right-hand panel]{vanderburg2020GOGREEN} reveals similarly good agreement at  $M_\star \gtrsim 10^{10.5}\,\msun$. With respect to GCLASS, the high mass end of the predicted GSMF is slightly too high, although still consistent with the observed relation within its $1\sigma$ error (grey shaded band). The agreement with GOGREEN at the high-mass end is essentially perfect.

At the low-mass end ($M_\star \lesssim 10^{10.5}\,\msun$), there is a significant discrepancy between the simulation and observations, in that the predicted GSMF is too high by a factor of up to $\approx$2 compared to both GCLASS and GOGREEN. Furthermore, there is \emph{a clear qualitative difference between the shape of the simulated and observed Schechter functions at increasingly higher redshift}: whereas the observed one shows a strong transition towards a down-turn at low masses (positive slope, i.e. $\alpha > -1$), the low-mass end of the Schechter function predicted by Hydrangea is closer to a flat profile with slow evolution at $0.6\leq z \leq 1.3$. As noted above, the low-mass slope parameter $\alpha$ predicted by Hydrangea also increases towards higher redshift, corresponding to a (relative) decrease of the GSMF (reflected by a 14 per cent increase in $\alpha$ from $z = 0.6$ to $z = 1.3$) at the low-mass end. However, the redshift evolution is evidently much stronger in the real Universe, where $\alpha$ increases by 35 per cent from $z\approx 0.6$ to $z\approx 1.3$. {\modtext This may hint at a failure in the simulation at disrupting low-mass satellites at $z \sim 1$ \citep[see][]{Bahe_et_al_2019} or overly efficient star formation in low-mass galaxies (e.g.~\citealt{Furlong2015eaglesmf} also find a slightly too high field SMF at $M_\star < 10^{10}\,\textrm{M}_{\odot}$ at these redshifts in EAGLE, see their fig. 2).

In principle, the discrepancy could also point to unaccounted systematic biases in the observations. While we note that multiple other studies have found a similar downturn (and flat slope) of the SMF at the low mass end \citep[e.g., ][]{Annunziatella2014,Nantais2016,Papovich_2018}, these studies may conceivably all be biased in a similar way; below, we therefore discuss in detail why we are confident that this is not the case before proceeding.

A first concern may be incompleteness: although the GCLASS and GOGREEN analyses of \citet{van2013environmental} and \citet{vanderburg2020GOGREEN} took particular care to account for this effect, the image simulations they used relied on assumptions about the properties of galaxies that were injected in these tests. Conceivably, a sub-population of very extended galaxies might therefore be missed in the data, since their low surface brightness makes them particularly difficult to observe. This would, however, require cluster galaxies to be biased to substantially larger sizes compared to the field, in contrast with current observational evidence (see e.g.~\citealt{Matharu_et_al_2019} and references therein). Secondly, observational biases may arise from the membership correction, i.e. the separation of galaxies into true cluster members and fore- / background objects. However, the GCLASS and GOGREEN analyses correct for this via  spectroscopic data sets (see \citealt{vanderburg2020GOGREEN}) and conclude from extensive tests that their results are robust. A third source of bias might be the stellar masses measurement of individual galaxies from SED fits, which require assumptions on the star formation history and the initial mass function (IMF). Differences in the assumed IMF in particular can affect the inferred stellar mass by a factor of $\approx$2 \citep{Marchesini2009ApJ}. However, since the simulations broadly match the observed field SMF at these redshifts \citep{Furlong2015eaglesmf}, invoking this explanation would require a very strong dependence of the IMF on environment, an assumption for which there is currently neither observational nor theoretical support. In short, none of these options is likely to yield a bias in the observations at the level of the discrepancy evident from Fig.~\ref{fig:modelSMFz06}.}

Coming back to the simulations, the low-mass slope of the the field SMF at similar redshift, as reported by \citet{Furlong2015eaglesmf}, is significantly steeper than what we find in the Hydrangea clusters: at $z = 0.5$ (2.0), \citet{Furlong2015eaglesmf} found a best-fit Schechter parameter $\alpha = -1.45$ (-1.57), compared to $\alpha = -1.06\ (-0.99)$ in the Hydrangea clusters. In addition, we note that the field SMF evolves towards a slightly steeper low-mass slope (more negative $\alpha$) with increasing redshift (see table A1 of \citealt{Furlong2015eaglesmf}), whereas the cluster environment (and interestingly also the field observations of \citealt{Muzzin_et_al_2013}) evolve towards a flatter or even inverted slope (less negative $\alpha$). Both differences indicate that, at the intermediate redshifts we probe, the cluster environment has a substantial impact on the SMF and hence allows an additional, and almost independent, test of the simulation model.

\section{Radial satellite density profiles within the clusters}
\label{denprofsec}

\label{nfwdis}
In the previous sections, we have demonstrated that the Hydrangea clusters successfully reproduce the scaling between the observed total stellar mass fractions and cluster mass (Sec.~\ref{sm_to_hm}), as well as the GSMF down to $\approx$~$3 \times 10^{10}\, \msun$ up to $z\approx1.5$, and down to at least $\approx$~$6\times10^9\,\msun$ at $z \approx 0.6$ (Sec.~\ref{smfdis}). Therefore, we now investigate the radial distribution of cluster satellites over time, comparing it to both observations and to the corresponding evolution of the dark matter (DM) distribution. As well as testing the simulations in an additional way that is orthogonal to the GSMF, we also aim to shed light on the previously reported difference between the observed increase in satellite concentration with redshift and the predicted decrease of the DM halo concentration \citep[][see the Introduction]{van2015evidence}.  

The density profile of relaxed DM haloes in N-body simulations, as well as the DM component of haloes in hydrodynamical simulations \citep{schaller2015effect} are robustly described by the NFW profile \citep[][see also \citealt{Dubinski_Carlberg_1991}]{navarro1997universal}, which has the form
\begin{equation} \label{nfweqn}
\frac{\rho(r)}{\rho_{cr}} = \frac{\delta_c}{\frac{r}{r_s}\left(1 + \frac{r}{r_s}\right)^2}.
\end{equation}
\noindent Here, $\rho_\mathrm{cr}$ is the critical density of the Universe, while $\delta_c$ and $r_s$ are the characteristic amplitude parameter and scale length of the profile, respectively, that can alternatively be expressed in terms of the halo radius $r_\mathrm{200c}$ and concentration parameter $c = r_\mathrm{200c}/r_s$. In the following analysis, we use the NFW profile in Eqn.~\ref{nfweqn} to model the density profiles of both the DM and of the stars contained in the Hydrangea cluster satellites\footnote{Recall that we are measuring the profile of satellites around the cluster, weighted by their stellar mass, rather than the distribution of stars within individual satellites.}, and carefully compare them with the observational results of \citet{van2015evidence}. For simplicity, we will generally refer to the NFW concentration parameter $c$ as ``the concentration'' for the remainder of this Section. 

\subsection{Evolution of the DM halo concentration}
\label{dmden}
As an initial check, we derive the evolution of the DM halo concentration from the 3D distribution of individual DM particles. For this, we calculate the density of DM particles in 20 consecutive shells, logarithmically spaced between 0.01 and 1.0 $r_\mathrm{200c}$ and centred on the cluster potential minimum\footnote{This radial range extends closer to the cluster centre than in e.g.~\citealt{neto2007statistics}, which is possible due to the higher resolution of our simulations. \citet{Gao_et_al_2008} have shown that this leads to a $\lesssim$10 per cent higher best-fit concentration, which is of no significance to our results.}. We include all DM particles within a given shell, irrespective of their subhalo membership. The resulting (3D) density profiles are fit with the NFW functional form (eq. \ref{nfweqn}) to obtain the characteristic radius $r_s$ and hence the concentration $c$, following e.g.  \citet{neto2007statistics} and \citet{schaller2015effect}. We repeat this process for each of the 24 Hydrangea clusters at 15 snapshots in the range $0 \leq z < 2$, and from these calculate the average concentration of the cluster ensemble and its confidence interval at each redshift.

\begin{figure}
	\includegraphics[width=\columnwidth]{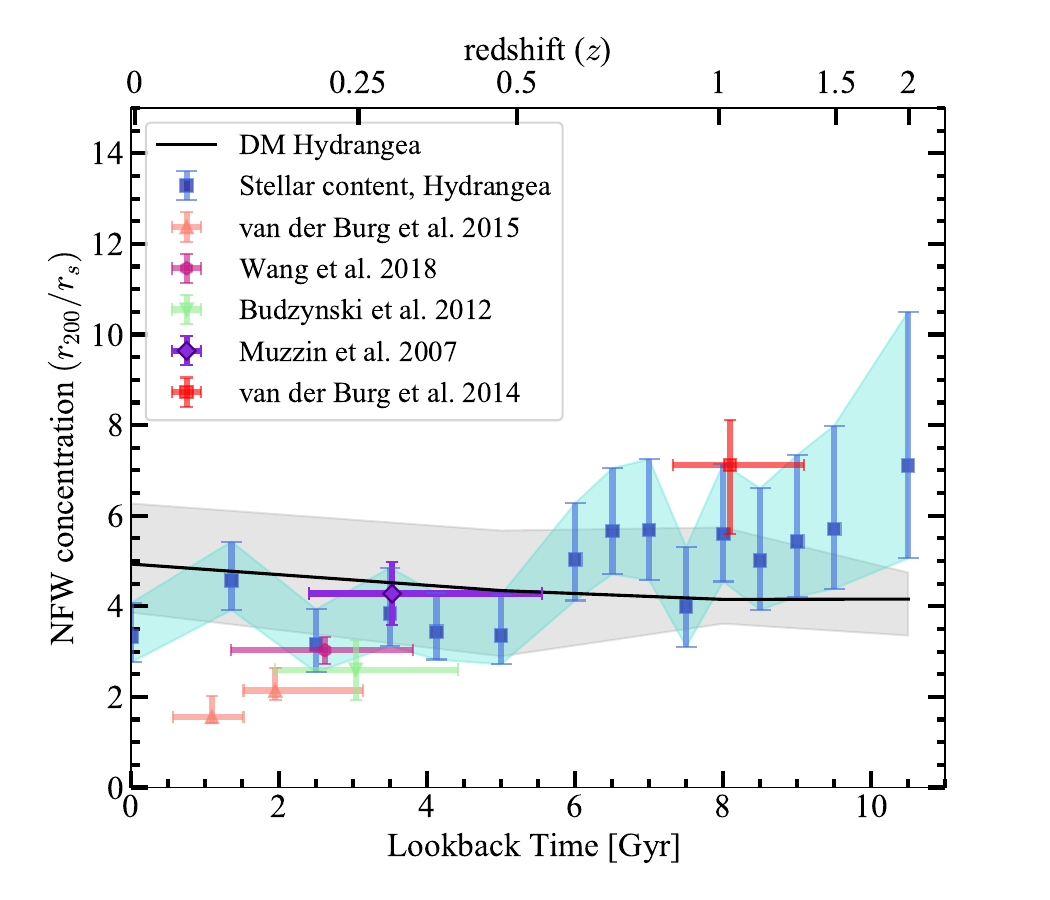}
	\caption{The redshift evolution of the NFW concentration of the dark matter (DM) halo and satellite galaxies for the Hydrangea clusters (black line and blue squares, respectively, with the shaded band and error bars enclosing regions with fitting $\chi^2 \leq 1$).The light-orange, light-green, magenta, purple, and red symbols represent corresponding observational measurements of the satellite galaxy concentrations from \citet{Muzzin_2007}, \citet{Budzynski2012MNRAS}, \citet{vdb14}, \citet{van2015evidence}, and \citet{wang2018satellite}, respectively. At $z \geq 0.25$, the simulations predict the same increase in galaxy concentrations with redshift as observed, while the DM halo concentration decreases. At low $z$, however, the simulated galaxy distribution is more strongly concentrated, by a factor of $\approx$2.} 
    \label{fig:nfw_evolution}
\end{figure}

The result is shown by the black line and grey shaded region in Fig.~\ref{fig:nfw_evolution}. In comparison with previous works \citep[e.g.][]{springel2005cosmological,neto2007statistics,duffy2008dark,dutton2014cold}, we find a qualitatively similar evolution of the DM halo concentration, decreasing by $\approx$20 per cent from $z = 0$ to $z = 2$. At a given redshift, the concentrations we measure are slightly higher than reported in some earlier studies, as expected from the difference in cosmological parameters \citep{dutton2014cold}.

\subsection{Concentration of stellar density profile}
We now analyze the corresponding galaxy density profile and its evolution. For consistency with observations, we use 2D profiles here, which we obtain by projecting the galaxy coordinates (relative to the cluster centre) along each of the three principal coordinate axes of the simulation; each galaxy is weighted by its stellar mass (which we measure within 30 pkpc, see Sec.~\ref{sec:mstar_subfind}) to yield a 2D galaxy density profile. {\modtext With the exception of the cluster BCG, we include all galaxies with stellar mass $M_\star > 10^9\, \msun$ \footnote{ \modtext This stellar mass limit is chosen to match that of \citet{van2015evidence}. At higher redshift, the data from \citet{vdb14} have a stellar mass limit of $\mstar \geq 10^{10.2}\, \msun$, but we have verified that this difference affects the best-fit NFW concentration only at a level $\ll 1\sigma$.}}within a cylinder of radius $2\times R_{200c}$ and (total) length along the projection axis equal to $4\times R_{200c}$ (we have verified that our results are insensitive to the choice of cylinder length\footnote{In the central region ($R_\mathrm{2D} \leq 0.5 \times R_\mathrm{200c}$), the density increases by $<$1 per cent if the cylinder length is increased further, from 4 to $6 \times R_\mathrm{200c}$.} and minimum $M_*$ in range $10^8$ -- $10^9 \mathrm{M}_{\odot}$). We then calculate the surface density within concentric annuli whose edge radii are spread linearly between $0.01$ and $0.1 \times R_{200c}$ ($\Delta R/R_\mathrm{200c} = 0.03$), and logarithmically between $0.1$ and $2 \times R_\mathrm{200c}$ ($\Delta \log R = 0.19$).

An important subtlety in this procedure is the way in which fore- and background galaxies are accounted for. Owing to the zoom-in nature of the Hydrangea simulations, our projected profiles cannot contain a contribution from galaxies at very large line-of-sight distances from the cluster, but correlated line-of-sight structures are still captured by the large Hydrangea zoom-in regions. For approximate consistency with the observational approach, we estimate and subtract the contribution of these nearby fore-/background galaxies to the projected galaxy density, rather than e.g. applying a cut in 3D radius or FoF membership. For this, we calculate the expected field galaxy density in each bin based on the galaxy volume density of the EAGLE S15\_AGNdT9-L0050N0752 simulation \citep{schaye2014eagle}, and subtract this from the surface density we measure for the Hydrangea clusters. This correction does not make any significant difference to the surface densities ($<0.17\%$ at all radii up to $R_{200\textrm{c}}$, and $<1\%$ at $R_{200\textrm{c}}<R<2 \times R_{200\textrm{c}}$), and does not affect the shapes of the density profiles.

We use EAGLE S15\_AGNdT9-L0050N0752 to derive the field galaxy density, rather than a volume near the edge of the Hydrangea simulations, because we have found that even near the edge of the zoom-in regions, the GSMF still has a slightly higher normalization (see Appendix \ref{sec:app_b}). This is not unexpected: clusters sit at the centre of large-scale overdensities in the cosmic web, with many galaxies merging into smaller groups before falling into the cluster halo instead of falling into the cluster halo individually from the field \citep[e.g.][]{McGee_et_al_2009}.

\begin{figure}
	\includegraphics[width=\columnwidth]{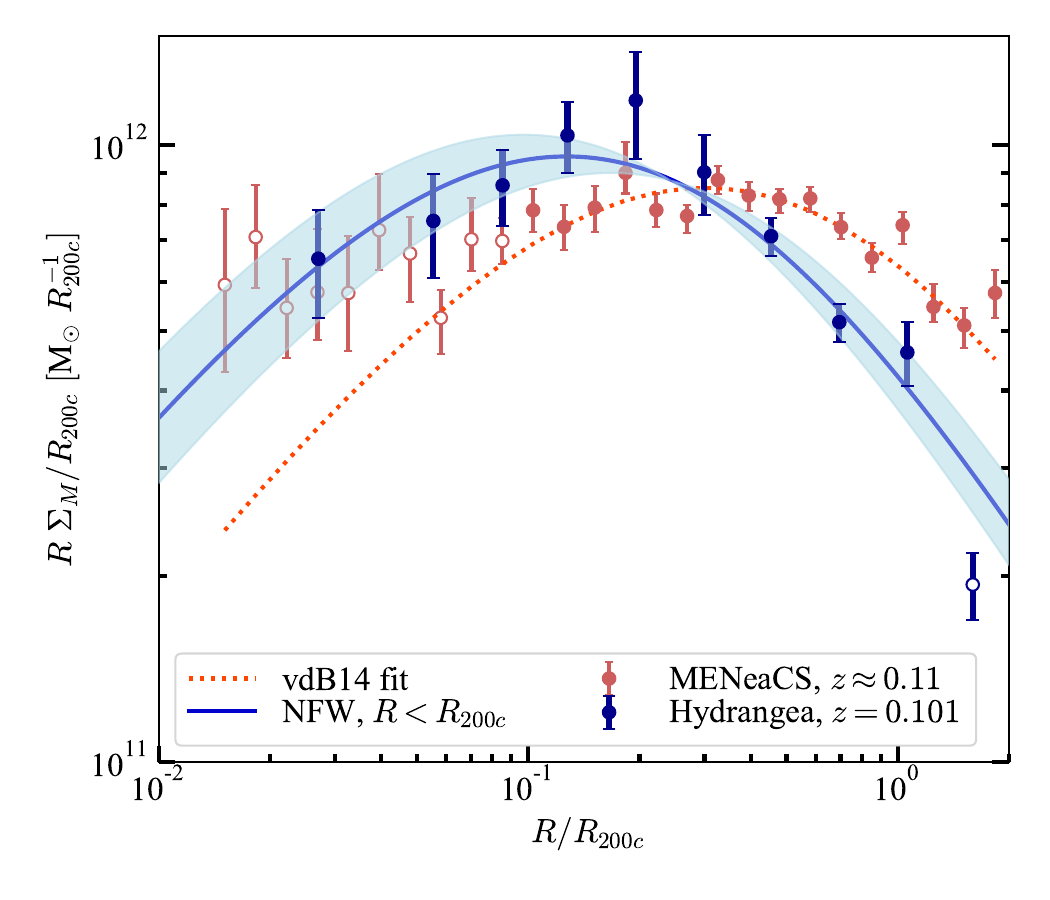}
	\caption{The stellar mass density distribution of the Hydrangea cluster ensemble at $z=0.101$. The dark-blue circles are the average surface densities at corresponding radial bins of the ensemble of 24 clusters. The dark-blue error bars are obtained from 100 bootstrap re-samplings with replacement from the full stack of galaxies. The blue solid line is the best-fit 2D projected NFW profile within $R_\mathrm{200c}$ and the shaded light-blue region shows the $\Delta \chi^2 = 1$ region for the fitted concentration parameter. Dark-orange circles represent the MENeaCS GSMF from \citet{van2015evidence} with their best-fit NFW profile shown as the dotted red line; both have been renormalized by the ratio of mean cluster masses in MENeaCS and Hydrangea (= 1.4) to enable an unbiased comparison. Filled circles represent bins included in the fit (for both the simulated and observed data), while open ones were excluded.}
	\label{fig:density_profile_0.101}
\end{figure}

To reduce statistical noise, we stack the individual density profiles (in units of $R / R_\mathrm{200c}$) from the 24 clusters at each redshift to obtain the average density profile for the simulation suite. As an example, Fig.~\ref{fig:density_profile_0.101} shows the density profile at $z = 0.101$ as dark-blue circles. 1$\sigma$ uncertainties on the surface density in each bin, shown by dark-blue error bars, are obtained from 100 bootstrap re-samplings of galaxies (from the full stack) with replacement, with the total number drawn in each instance equal to the total number of galaxies in the stack. To reduce the dynamic range, we show the profile multiplied by $R / R_\mathrm{200c}$. In this form, it shows  -- as expected for an NFW profile -- a marked downturn at $R \gtrsim 0.2 R_\mathrm{200c}$, and a similar, but weaker, drop towards lower $R$.


We fit each stacked density profile with the 2D projected NFW profile \citep{bartelmann1996arcs},
 \begin{equation} \label{2dnfw}
 \Sigma (x) = \frac{2\rho_s r_s}{x^2 - 1} f(x)
 \end{equation}
 
 \noindent where $x \equiv r/r_s$ and
\[
f(x)= 
\begin{cases}
1 - \frac{2}{\sqrt{x^2-1}} \textrm{arctan} \sqrt{\frac{x-1}{x+1}},&  (x\geq 1)\\
1 - \frac{2}{\sqrt{1 -x^2}} \textrm{arctanh} \sqrt{\frac{1-x}{1 +x}},&  (x\leq 1)\\
0,              & (x=1),
\end{cases}
\]
to obtain the concentration parameter, minimizing the $\chi^2$ value between the functional form and the densities in each bin. For the $z \approx 0.1$ example shown in Fig.~\ref{fig:density_profile_0.101}, the best-fit NFW profile is shown as a solid blue line, with the light blue shaded band marking the region that is covered by $\Delta\chi^2 \leq 1$ values of the fit. Within the uncertainties, this fit provides a good description of the density profile.

A point worth mentioning here is that the NFW profile is fitted only within a (projected) radius of 1, rather than $2 \times R_\mathrm{200c}$, from the cluster centre of potential as in most of the observational studies that we compare to. As can be seen in Fig.~\ref{fig:density_profile_0.101}, the rightmost dark-blue point (open circle) at $R > R_\mathrm{200c}$ lies far below the best-fit NFW profile. Including this point in the fitting procedure would drive up the best-fit concentration and force a shift of the blue line towards the right side of the figure, resulting in an overall much worse fit (both visually and in terms of the best-fit $\chi^2$ value). This behaviour is present in all the snapshots within the redshift range $0< z <2$. Therefore, we conclude that the stellar surface density profile of the Hydrangea clusters is well represented by the NFW profile only up to a distance of $R_\mathrm{200c}$ from the cluster centre, and use this radial distance limit in our fitting procedure for all the redshifts. {\modtext This changed radial limit compared to the observations may lead to a slight difference in the estimated value of the concentration. However, as the concentration of the density profile is mainly dependant on the characteristic radius $r_s$ in terms of including the radial range, having extended profiles beyond $r_{200}$ does not make any significant difference to the estimation for the concentration range we are concerned about in this work ($r_s \leq 0.7 \times r_{200}$ for $c \geq 1.5$).}

At each redshift, we calculate the best-fit satellite concentration for three orthogonal projections as described above. Figure~\ref{fig:nfw_evolution} shows the evolution of the mean satellite concentration (averaged over all three projections) over redshift as blue points, with error bars representing the standard deviation between projections. The area within the error bars is shaded light blue for clarity. The light-orange, light-green, magenta, purple, and red symbols show observationally measured satellite concentrations over different redshift ranges, as indicated by the horizontal error bar \citep{Muzzin_2007,Budzynski2012MNRAS,vdb14,van2015evidence,wang2018satellite}.

Compared to the DM concentration (black line and gray shaded region), the satellite concentration clearly shows a different trend with redshift, especially at $z\geq 0.5$ where they are higher by up to a factor of 2. The predicted satellite concentrations agree well with what is observed at both $z \approx 1$ \citep[][red circle]{vdb14} and $z \approx 0.3$ \citep[][purple diamond]{Muzzin_2007}, strongly suggesting that this is difference is not an artefact of the simulation. At lower redshift, $z < 0.5$, the satellite concentration in Hydrangea closely traces that of the DM halo. This is in marked contrast to the MENeaCS data of \citet{van2015evidence}, which indicate a continued decrease in satellite concentration with time, down to $c \lesssim 2$ at $z = 0$: at the present day, the Hydrangea concentrations are more than a factor of two higher than what is observed.

\subsection{Interpreting the concentration evolution of DM and satellites}

\citet{van2015evidence} interpreted their finding of decreasing satellite concentration with time as evidence for an inside-out growth of the satellite halo of massive clusters: at high redshift, satellites are strongly concentrated towards the cluster centre, but over time new satellites are preferentially added to the cluster outskirts while tidal stripping reduces the mass (or even number) of satellites at small cluster-centric radii (see also \citealt{Bahe_et_al_2019}). The excellent match between the satellite concentrations predicted by Hydrangea and inferred from observations at redshifts $z \gtrsim 0.25$ supports this picture: high-redshift satellite haloes are indeed more concentrated than the diffuse DM haloes built up at the same time. \emph{The observational finding that the satellite concentrations are higher than DM haloes from $N$-body simulations is therefore a consistent prediction of $\Lambda$CDM}, rather than an indicator for incompleteness in the high-redshift observations or tensions with the cosmological model. 

At lower redshift, however, the Hydrangea simulations are clearly in disagreement with observations on the satellite halo profile: the predicted concentrations are too high by more than a factor of two. The seemingly unchanged concentration at $z \lesssim 0.5$ in the simulations suggests almost self-similar galaxy cluster growth at all radial distances, in contrast to the inside-out scenario indicated by the observations. This tension is somewhat surprising: given that the EAGLE simulation model was calibrated against observations of the local, rather than high-redshift, Universe, one might instead have expected the simulation to match the observed satellite concentrations better at lower redshifts. 

To investigate this discrepancy further, we plot the observed satellite density profile from the MENeaCS clusters \citep{van2015evidence} in Fig.~\ref{fig:density_profile_0.101} as dark-orange points, with the dotted red line giving the best-fit projected NFW profile as found by these authors. To account for the slightly mass offset between MENeaCS and Hydrangea, these observed densities have been scaled by a factor of 1.4, the ratio of the mean cluster masses in MENeaCS and Hydrangea at $z = 0.1$ (see Sec.~\ref{menobs}). \citet{van2015evidence} reported an over-dense central region within $0.1\times R_\mathrm{200c}$ that can not be fitted with an NFW profile (shown with dark-orange open circles), and therefore restricted their fit to the range $0.1<R/R_\mathrm{200c}<2.0$ where their profile is well described by an NFW profile with concentration parameter $c = 2.03\pm 0.2$ (data points shown by filled grey circles in Fig.~\ref{fig:density_profile_0.101}). The corresponding best-fit Hydrangea profile (blue line in Fig.~\ref{fig:density_profile_0.101}, with a concentration parameter of $c = 4.5\pm 1.01$) likewise provides a good fit to the simulated satellite profile.

Comparing the individual bin densities, the simulations predict an approximately realistic satellite density 
within $\approx$0.1 $R_\mathrm{200c}$ of the cluster centre (the simulated densities are $\approx$15 per cent higher, but this is still within the uncertainties). A more significant excess ($\approx$30 per cent) is predicted around $\approx$0.2 $R_\mathrm{200c}$, followed by a similarly large shortfall at $R_\mathrm{200c}$. To see whether the Hydrangea and MENeaCS satellite density profiles are compatible in terms of their total stellar mass, we have integrated them over the range $0.1<R/R_{200c}<1.0$. With the normalization correction as described above, the two profiles are within 10 per cent of each other in terms of total mass (slightly higher for MENeaCS). Consistent with our earlier results (Section \ref{sm_to_hm}), this confirms that the total stellar mass of the low-$z$ Hydrangea satellite haloes is realistic, while its radial distribution is not. 

While the low concentration of cluster satellite haloes in the local Universe is observationally well-established \citep{Lin2004ApJ,Budzynski2012MNRAS,van2015evidence}, the picture is less clear on the theoretical side. In a semi-analytic model of galaxy formation, \citet{Wang_et_al_2014} have found that the distribution of satellites around massive isolated galaxies traces almost exactly that of the DM halo, as is the case in Hydrangea. These authors speculate that poor modelling of tidal disruption in their model may be the reason for the overly steep satellite halo profile, but this process is accounted for self-consistently in hydrodynamical simulations such as Hydrangea. In the BAHAMAS simulation\footnote{Like Hydrangea, this is a cosmological hydrodynamic simulation, but it models a large volume at much lower resolution.}, \citet{McCarthy2017Bahama} obtain a best-fit concentration of $c \approx 1.8$, in close agreement with the observations of \citet{van2015evidence}. In detail, however, their (3D) profile also shows a slight ($\approx$10 per cent) excess compared to the observations at $r/r_\mathrm{200c} \approx 0.3$, and follows a DM-like profile with $c = 4$ at larger radii (their fig.~9).

It is therefore unclear at present whether the tension in the low-redshift satellite halo concentrations is a particular shortcoming of Hydrangea, or may be a more generic deficiency of cosmological simulations; further work analysing the predictions from other recent theoretical models is necessary to establish a clearer picture. Two factors that might bias the satellite halo concentration high in Hydrangea are the selection of the clusters to be relatively isolated at low redshift -- which might, conceivably, increase the fraction of clusters that have undergone significant merging shortly before $z = 0$, and have not yet reached an equilibrium state -- and their overly massive BCGs \citep{bahe2017hydrangea}. The latter shortcoming, which is similarly affecting other simulations \citep[e.g.][]{pillepich2018illustris}, may influence the dynamics, and hence density, of the central satellite halo, although we found in Fig.~\ref{fig:density_profile_0.101} that the differences between Hydrangea and the MENeaCS data of \citet{van2015evidence} primarily originate in the cluster outskirts. More work is necessary to establish whether either of these aspects is related to the overly concentrated profile of satellite haloes in Hydrangea clusters.

\section{Summary and Conclusions} 
\label{sumandconc}
Observations have shown that the high-redshift satellite galaxy haloes of massive clusters were significantly more concentrated than in the present Universe, while DM halo concentrations are predicted to evolve in the opposite direction. To gain insight into the nature of this evolution, we have analyzed the stellar content, galaxy stellar mass function, and satellite halo concentration in the Hydrangea suite of 24 massive galaxy clusters ($M_\mathrm{200c} >10^{14}\, \textrm{M}_{\odot}$ at $z=0$) over the redshift range $0 < z < 2$. From this analysis, including comparison to observational data up to $z \approx 1.3$, we draw the following conclusions: 

\begin{enumerate}

    \item Stellar masses of simulated galaxies, as measured with the \textsc{\subfind{}} halo finder and within an aperture of 30 pkpc, agree well with masses recovered from synthetic images with \textsc{SExtractor}; they can therefore be compared meaningfully to stellar masses obtained from observational data (Fig.~\ref{synismfmg}).
    
    \item The scaling relation of the cluster stellar mass fraction with respect to the cluster mass predicted by the Hydrangea simulations at $0.1<z<1.4$ agrees well with the equivalent relation reported from observational data \citep{chiu2018}: the offsets in the best-fit power law index (1 per cent) and normalization (12.5 per cent) are (well) within $1\sigma$ uncertainties. The total stellar content of Hydrangea clusters is therefore realistic out to at least $z = 1.4$ (Fig.~\ref{fig:chiuftest}).
    
   \item The galaxy stellar mass functions of Hydrangea clusters are well described by Schechter functions, whose best-fit normalization $\Phi^*$ depends approximately linearly on cluster mass, whereas the characteristic mass $\mathscr{M}^*$ and low-mass slope $\alpha$ show no significant trend with mass (Fig.~\ref{fig:schechterparams}). From this $\Phi^*$--$M_\mathrm{cluster}$ relation, we have constructed a scaled Schechter function to compare the Hydrangea predictions to observations of clusters with moderately different masses (Table~\ref{tab:model_params}).
    
    \item Accounting for differences in the cluster mass distribution, the satellite GSMF predicted by Hydrangea shows an excellent match to observations up to $z = 1.3$, at $M_\star > 10^{10.5}\,\msun$. At $M_\star < 10^{10.5}\,\msun$, the predicted GSMF from Hydrangea is higher than observed by up to a factor of $\approx$2 and the simulations do not reproduce the qualitative shift from a negative to positive low-mass slope that is seen in observations. The best-fit $\alpha$ parameter also increases in the simulations, but only by 14 per cent, as opposed to a 35 per cent increase in the observations (Fig.~\ref{fig:modelSMFz06}).
        
    \item The concentration of the satellite halo profile (weighted by stellar mass) in Hydrangea clusters is up to a factor of 2 higher than that of the DM halo at redshift $z \gtrsim 0.5$, in agreement with observations. This discrepancy therefore stems from differences in the assembly of the satellite and DM halo; the high satellite concentrations are fully consistent with expectations from $\Lambda$CDM (Fig.~\ref{fig:nfw_evolution}).

    \item At low redshift ($z \lesssim 0.3$), the satellite halo concentration in Hydrangea clusters closely matches that of the DM halo and is too high by a factor of $\approx$2 at $z = 0$ compared to observations (Fig.~\ref{fig:nfw_evolution}). From a direct comparison of the full satellite halo profile to observations at $z = 0.1$, the simulations predict a satellite halo density that is too high in the inner part ($\approx$0.2 $R_\mathrm{200c}$) and too low at larger radii ($\gtrsim$0.3 $R_\mathrm{200c}$), both by factors of $\approx$30 per cent (Fig.~\ref{fig:density_profile_0.101}). The physical origin of this inaccuracy, which has also been reported from a semi-analytic galaxy formation model \citep{Wang_et_al_2014}, remains unclear. 
    
\end{enumerate}

Our analysis confirms that the Hydrangea simulations, and the underlying EAGLE simulation model, predict broadly realistic stellar properties of galaxy clusters, not just at low redshift \citep{bahe2017hydrangea}, but also across more than half the history of the Universe. They can therefore be used to make meaningful predictions about the transformation of cluster galaxies, and the environmental mechanisms that cause them. At the same time, there are clear deviations from the real Universe in detail: at high redshift the simulations contain too many low-mass galaxies, and at low redshift the satellite halo is too concentrated. Both deficiencies may be related to overly efficient star formation in the simulations, or alternatively due to stellar stripping being suppressed by the limited resolution or as a consequence of too high stellar concentration within satellites. Future simulations can use these diagnostics to improve the fidelity of galaxy formation modelling.

\section*{Acknowledgements}
{\modtext We thank the referee for valuable comments that helped to improve the presentation of our results.} The authors acknowledge support from the Netherlands Organization for Scientific Research (NWO) under Vici grant number 639.043.512 (SLA, HH), and Veni grant number 639.041.751 (YMB). YMB also acknowledges funding from the EU Horizon 2020 research and innovation programme under Marie Sk\l odowska-Curie grant agreement 747645 (ClusterGal). The Hydrangea simulations were in part performed on the German federal maximum performance computer ``HazelHen'' at the maximum performance computing centre Stuttgart (HLRS), under project GCS-HYDA / ID 44067 financed through the large-scale project ``Hydrangea'' of the Gauss Center for Supercomputing. Further simulations were performed at the Max Planck Computing and Data Facility in Garching, Germany. This work used the DiRAC@Durham facility managed by the Institute for Computational Cosmology on behalf of the STFC DiRAC HPC Facility (www.dirac.ac.uk). The equipment was funded by BEIS capital funding via STFC capital grants ST/K00042X/1, ST/P002293/1, ST/R002371/1 and ST/S002502/1, Durham University and STFC operations grant ST/R000832/1. DiRAC is part of the National e-Infrastructure.

This work has made use of Python (http://www.python.org), including the packages \textbf{numpy} \citep{Harris_et_al_2020} and \textbf{scipy}
(http://www.scipy.org). Plots have been produced with \textbf{matplotlib} \citep{hunter2007matplotlib}. 


\section*{Data Availability}
The data presented in the figures are available upon request from the corresponding author. The raw simulation data can be requested from the C-EAGLE team \citep{bahe2017hydrangea,barnes2017cluster}.



\bibliographystyle{mnras}
\bibliography{SMFDP_Hydrangea} 



\appendix
{\modtext \section{Testing different noise levels for SMF recovery from synthetic images}
\label{app:noise_lev_test}

In Sec.~\ref{sec:mstar_subfind}, we verified that the \subfind{} subhalo catalogues from the simulation output yield satellite stellar masses that are compatible with those derived by \textsc{SExtractor} from synthetic images. In Fig.~\ref{synismfmg}, we showed the comparison of the SMF obtained from \subfind{} and \textsc{SExtractor} run on synthetic images that include noise at an RMS level of $1.5\times 10^6$ M$_{\odot}$ per pixel. In Fig.~\ref{fig:app_noise_level}, we show the analogous comparison to the SMF derived from \subfind{} (blue) and synthetic images (magenta) with three different RMS noise levels:  $7.5\times 10^5$ (left), $3.0\times 10^6$ (middle), and $6.0\times 10^6\,\msun\,\mathrm{pixel}^{-1}$ (right); except for the different noise levels, these are produced in exactly the same way as described in Sec.~\ref{sec:mstar_subfind}. Error bars represent the Poisson errors obtained from 100 bootstrap re-samplings of the stack of galaxies in each sample. In each case, the stellar mass functions from \subfind{} and those from the synthetic images agree within their uncertainties. Likewise, the best-fit Schechter functions to the SMFs (solid lines) all agree well within the error bars with the one from \subfind{}.

On closer inspection, however, two subtle trends of the best-fit Schechter functions with respect to the injected noise level become detectable. Firstly, the best-fit low-mass slope parameter $\alpha$ from the \textsc{SExtractor} catalogues (magenta text) is systematically lower than its \subfind{} equivalent (blue text), at all noise levels, even though the discrepancy is only at the $\approx$1$\sigma$ level. This shift in $\alpha$ is mainly caused by projections at the lower mass end. As the \textsc{SExtractor} output gives the stellar masses of all the diffuse line-of-sight star particles from the simulation region along the direction of the projection, it increases the total stellar mass of each source by a small value even if there is only one source along the line of sight. This increase in the stellar mass is more significant for galaxies with a stellar mass lower than $10^9\ \textrm{M}_{\odot}$. We have verified that this is indeed the explanation by comparing the stellar masses of the detected and matched sources between \subfind{} and \textsc{SExtractor} outputs for all the stellar masses (not shown here).

Secondly, the best-fit characteristic mass $\mathscr{M}^*$ from \textsc{SExtractor} decreases systematically with increasing noise levels, while it is (unsurprisingly) constant for \subfind{}. Conceivably, this is because in noisier images, \textsc{SExtractor} recovers less stellar mass for the individual sources. This loss is more prominent at the high-mass end, where galaxies have more extended and diffuse stellar haloes that become lost in the noise. Because of this, galaxies with true stellar mass above $10^{10}\ \textrm{M}_{\odot}$ are assigned a lower stellar mass by \textsc{SExtractor}, lowering $\mathscr{M}^*$. To confirm and quantify the trend of lower $\mathscr{M}^*$ with noise levels, and resolve the apparent paradox that the best match to \subfind{} is seen for the highest noise level, we have fitted the \textsc{SExtractor} SMFs with Schechter functions again, but with $\alpha$ fixed to the \subfind{} average of $-1.02$ (not shown here). In this case, the lowest noise level yields a best-fit value of $\mathscr{M}^* = 10^{10.92}\,\textrm{M}_{\odot}$, very close to the \subfind{} value of $10^{10.94}\  \textrm{M}_{\odot}$); the value then decreases with increasing noise (by $\approx$0.1 dex at the highest noise level), consistent with the trends in Fig.~\ref{fig:app_noise_level}. 

The slight difference in the fitted $\alpha$ value of the SMF from the \subfind{} outputs is caused by the matching of detected sources with the \textsc{SExtractor} output. The change is $<<1 \sigma$ value of the fitted $\alpha$ parameter, and therefore, considered negligible in this case. 

From the close agreement between the simulations and the synthetic images for all the four different noise levels we have tested for, we conclude that the subhalo stellar mass within 30 pkpc as measured by the \subfind{} code can be reliably used for the analysis presented in the main text. However, the slight variation of the SMF shape with noise level may indicate that at higher redshifts, where the noise levels in the observed data are higher, our use of \subfind{} masses may cause a very small positive bias in the massive end of the GSMF.


\begin{figure*}	
	\includegraphics[width=1.99\columnwidth]{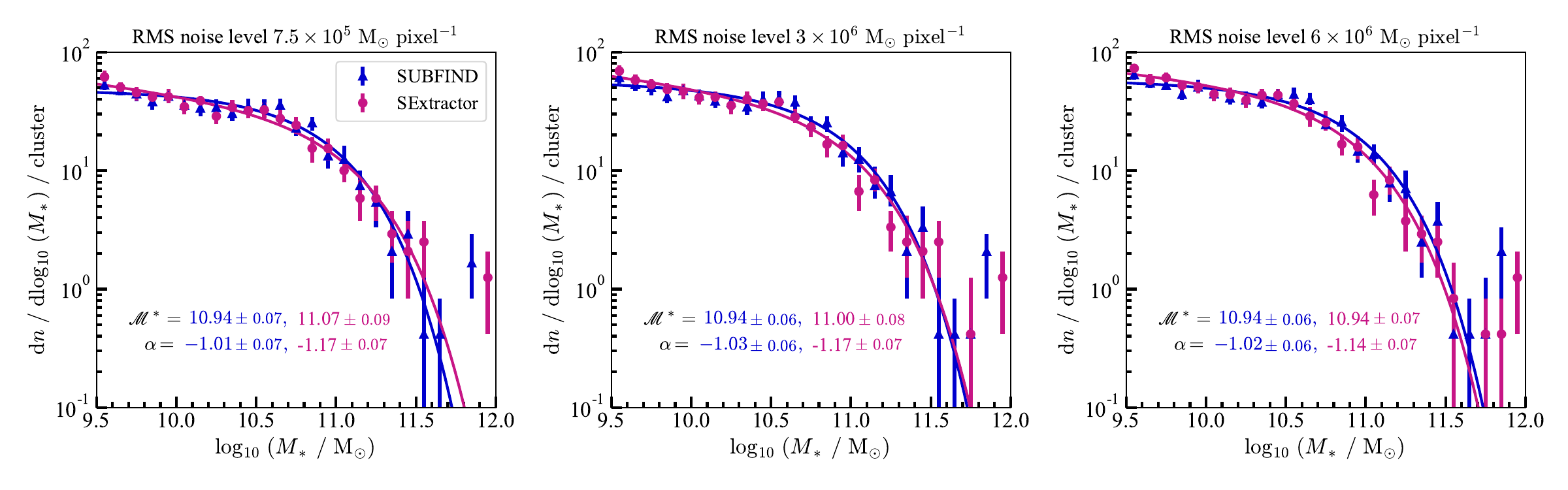}
	\caption{{\modtext The galaxy stellar mass function (GSMF) of the simulated clusters (coloured symbols) and their best-fit Schechter functions (solid lines) for RMS noise levels of $7.5\times 10^5$ (left), $3.0\times 10^6$ (middle), and $6.0\times 10^6\,\msun\,\mathrm{pixel}^{-1}$ (right). For all the panels, the blue points are obtained from the subhalo stellar mass measured by \subfind{} within 30 pkpc from the centre of potential of each subhalo; magenta points are obtained from the estimated stellar mass of each galaxy by running \textsc{SExtractor} on synthetic images (see Sec. \ref{sec:mstar_subfind} for details). The error bars indicate 1$\sigma$ uncertainties obtained from bootstrap re-samplings of the stack of galaxies in each sample. Both the mass functions and their Schechter fits agree within statistical uncertainties, indicating that the \subfind{} mass measurement is consistent with the observational approach. }}
	\label{fig:app_noise_level}
\end{figure*}
}


\section{Fitted Schechter function parameters}
In Table \ref{tab:fitted_params}, we list the best-fit Schechter parameters for the individual Hydrangea clusters, or stacks thereof, at $z \approx 0.6$. These data correspond to what is plotted in Fig.~\ref{fig:schechterparams} in the main paper. 

\begin{table}
	\centering
	\caption{Fitted parameters for the Schechter function of the simulated clusters at $z=0.6$. The left column indicates the cluster number from the Hydrangea suite. $M_{500c}$ indicates the corresponding cluster halo mass at that redshift. $\Phi^*$ is the overall normalization, $M^*$ is the characteristic mass, and $\alpha$ is the low-mass slope. The first four results are from groups of four low-mass clusters each with $M_h \leq 1.75 \times 10^{14}\textrm{M}_{\odot}$. The group ID in the first column is given with increasing average halo mass of the cluster groups, and the average halo mass is given in the second column with the individual cluster halo mass.}
	\label{tab:fitted_params}
	\begin{tabular}{lcccr} 
		\hline
		Cluster  & $M_{500c}$ & $\Phi^*$ & $\log_{10}[M^*/M_{\odot}]$ & $\alpha$\\
		ID & $[10^{14} M_{\odot}]$ \\
		\hline
		Cgr-1 & 0.51 & $10.29\pm4.24$ & $10.79\pm0.18$ & $-0.93\pm0.16$\\
		Cgr-2 & 0.78 & $10.16\pm4.07$ & $11.05\pm0.14$ & $-1.15\pm0.12$\\
		Cgr-3 & 1.02 & $17.68\pm5.75$ & $10.89\pm0.13$ & $-1.05\pm0.13$\\
		Cgr-4 & 1.38 & $14.78\pm4.46$ & $11.06\pm0.11$ & $-1.21\pm0.09$\\
		CE-12 & 2.39 & $42.20\pm19.54$ & $10.89\pm0.19$ & $-1.02\pm0.19$\\
		CE-18 & 2.49 & $35.15\pm14.59$ & $11.03\pm0.16$ & $-1.10\pm0.14$\\
		CE-22 & 4.19 & $69.62\pm22.71$ & $10.97\pm0.13$ & $-1.02\pm0.12$\\
		CE-24 & 3.52 & $48.75\pm20.10$ & $10.92\pm0.14$ & $-1.15\pm0.15$\\
		CE-25 & 2.32 & $72.26\pm25.41$ & $10.66\pm0.16$ & $-0.90\pm0.20$\\
		CE-28 & 2.57 & $28.75\pm9.62$ & $11.18\pm0.12$ & $-1.27\pm0.08$\\
		CE-29 & 5.21 & $96.38\pm31.04$ & $10.74\pm0.14$ & $-0.92\pm0.17$\\
		\hline
	\end{tabular}
\end{table}

\section{Comparison to the field environment}
\label{sec:app_b}
\begin{figure}
    \centering
    \includegraphics[width=\columnwidth]{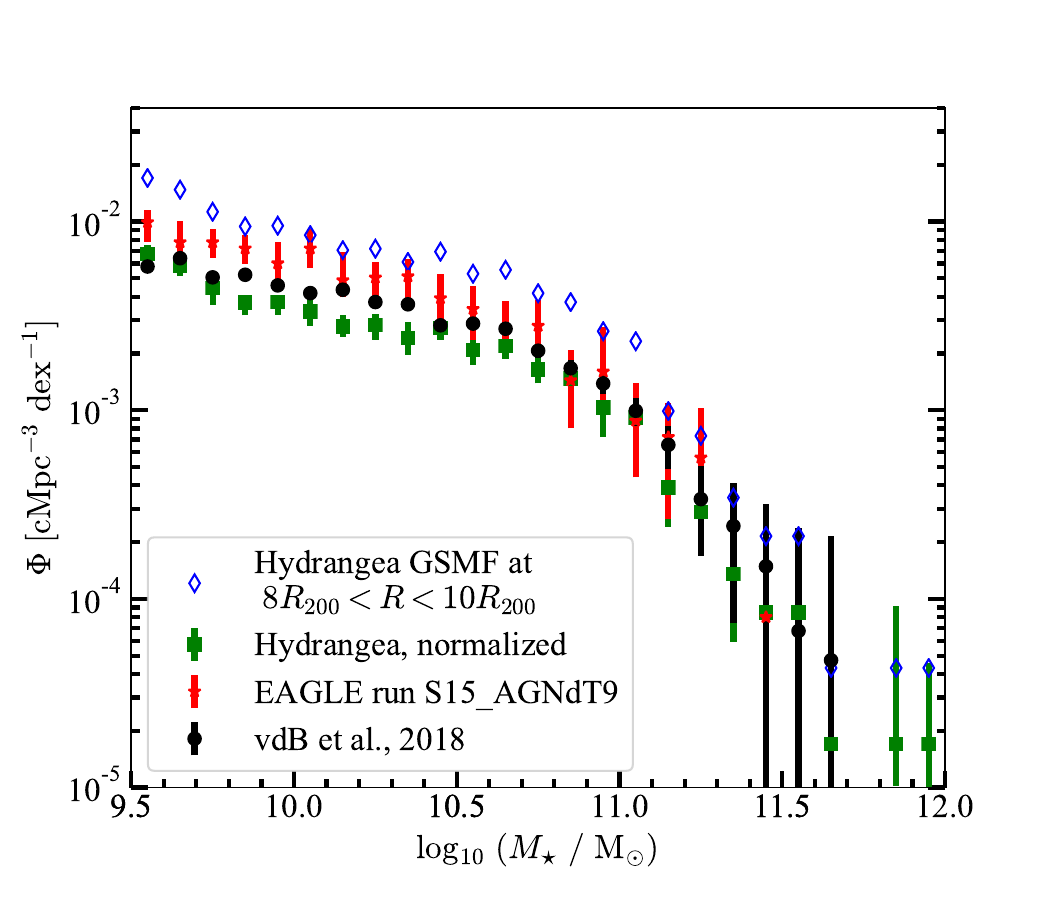}
    \caption{The galaxy stellar mass function (SMF) from simulations and observation at $z\approx 0.6$. The black circles show the SMF of the field environment from the observational data from COSMOS/UltraVISTA survey \citep{van2018stellar}. The open blue diamonds are the SMF from the Hydrangea simulations at a projected distance range $8 \times R_\mathrm{200c} < R < 10 \times R_\mathrm{200c}$ from the centre of potential of each central galaxy clusters. The green squares are the same SMF shown in the blue open diamonds, re-normalized with the average density of the Universe at this redshift. The red stars are SMF from the EAGLE run S15\_AGNdT9-L0050N0752 \citep{schaye2014eagle}.}
    \label{fig:field_approx}
\end{figure}

The outer regions of galaxy clusters are observed to have higher number density of galaxies compared to the average field environment. Therefore, before estimating the radial density profile of the clusters in Sec. \ref{dmden}, we subtract the average field density from the stellar density profile obtained from the Hydrangea clusters. 
To approximate the field environment, we test with the cluster SMF at different 2D projected annuli at increasing distances from the center of potential of the clusters ($3R_{200\textrm{c}}<R<5R_{200\textrm{c}}$ to $8R_{200\textrm{c}}<R<10R_{200\textrm{c}}$) and compare with field SMF from observational data at $z\approx 0.6$ from \citet{van2018stellar}. We observe a steady trend of the SMF having lower normalization at higher radial distance from the cluster center. However, even at the projected annulus between $8R_{200\textrm{c}}<R<10R_{200\textrm{c}}$, the cluster SMF has a higher normalisation factor compared to the observed field SMF as shown by the open blue diamonds (Hydrangea and $8R_{200\textrm{c}}<R<10R_{200\textrm{c}}$) and black circles (COSMOS/UltraVISTA field SMF from \citet{van2018stellar}) in Fig. \ref{fig:field_approx}. The field SMF matched within errorbars with the observations after re-normalizing the matter density in the simulated clusters with the average density of the universe (green squares in Fig. \ref{fig:field_approx}). To recheck, we compare the SMF with the field SMF from the EAGLE run S15\_AGNdT9-L0050N0752 \citep{schaye2014eagle} at $z\approx0.6$ and it matched with the observations (red stars in Fig. \ref{fig:field_approx}). This indicates that even at a large distance from the cluster center, the outskirts of the simulated clusters are over-dense compared to the average field environment.
 

\bsp	
\label{lastpage}
\end{document}